\documentclass[12pt,english]{extarticle}
\usepackage[T1]{fontenc}
\usepackage[latin9]{inputenc}
\usepackage[a4paper]{geometry}
\usepackage{color}
\usepackage{float}
\usepackage{url}
\usepackage{amsmath}
\usepackage{amsthm}
\usepackage{amssymb}
\usepackage{graphicx}
\usepackage{setspace}
\usepackage[authoryear]{natbib}
\makeatletter
\theoremstyle{plain}
\newtheorem{assumption}{\protect\assumptionname}
\theoremstyle{plain}
\newtheorem{lem}{\protect\lemmaname}
\theoremstyle{plain}
\newtheorem{thm}{\protect\theoremname}
\theoremstyle{plain}
\newtheorem{prop}{\protect\propositionname}

\@ifundefined{date}{}{\date{}}
\bibpunct{(}{)}{,}{a}{,}{,}

\usepackage{babel}

\usepackage{chngcntr}
\usepackage{apptools}
\AtAppendix{\counterwithin{lemma}{section}}

\usepackage{scrextend}

\makeatletter

\def\thmhead@plain#1#2#3{%
  \thmname{#1}\thmnumber{\@ifnotempty{#1}{ }\@upn{#2}}%
  \thmnote{ {\the\thm@notefont#3}}}
\let\thmhead\thmhead@plain

\def\th@remark{%
  \thm@headfont{\bfseries}%
  \normalfont 
  \thm@preskip\topsep \divide\thm@preskip\tw@
  \thm@postskip\thm@preskip
}

\makeatother

\usepackage{pgfpages}

\usepackage{setspace}

\newcommand{\mnorm}[1]{{\lvert\kern-0.25ex\lvert\kern-0.25ex\lvert #1 
    \rvert\kern-0.25ex\rvert\kern-0.25ex\rvert}}

\usepackage{arydshln}

\makeatother

\usepackage{babel}
\providecommand{\assumptionname}{Assumption}
\providecommand{\lemmaname}{Lemma}
\providecommand{\propositionname}{Proposition}
\providecommand{\theoremname}{Theorem}

\begin{document}
\title{\vspace{20pt}
Subgeometrically ergodic autoregressions with autoregressive conditional
heteroskedasticity\thanks{The authors thank the Academy of Finland (MM and PS), Foundation for
the Advancement of Finnish Securities Markets (MM), and OP Group Research
Foundation (MM) for financial support, and Co-Editor Robert Taylor
and three anonymous referees for useful comments and suggestions.
Contact addresses: Mika Meitz, Department of Economics, University
of Helsinki, P. O. Box 17, FI\textendash 00014 University of Helsinki,
Finland; e-mail: mika.meitz@helsinki.fi. Pentti Saikkonen, Department
of Mathematics and Statistics, University of Helsinki, P. O. Box 68,
FI\textendash 00014 University of Helsinki, Finland; e-mail: pentti.saikkonen@helsinki.fi.}\vspace{20pt}
}
\author{Mika Meitz\\\small{University of Helsinki} \and Pentti Saikkonen\\\small{University of Helsinki}\vspace{20pt}
}
\date{First version May 2022, revised April 2023}
\maketitle
\begin{abstract}
\noindent In this paper, we consider subgeometric (specifically,
polynomial) ergodicity of univariate nonlinear autoregressions with
autoregressive conditional heteroskedasticity (ARCH). The notion of
subgeometric ergodicity was introduced in the Markov chain literature
in 1980s and it means that the transition probability measures converge
to the stationary measure at a rate slower than geometric; this rate
is also closely related to the convergence rate of $\beta$-mixing
coefficients. While the existing literature on subgeometrically ergodic
autoregressions assumes a homoskedastic error term, this paper provides
an extension to the case of conditionally heteroskedastic ARCH-type
errors, considerably widening the scope of potential applications.
Specifically, we consider suitably defined higher-order nonlinear
autoregressions with possibly nonlinear ARCH errors and show that
they are, under appropriate conditions, subgeometrically ergodic at
a polynomial rate. An empirical example using energy sector volatility
index data illustrates the use of subgeometrically ergodic AR\textendash ARCH
models.

\bigskip{}
\bigskip{}

\noindent\textbf{JEL classification:} C22.

\bigskip{}

\noindent\textbf{MSC2020 classification}s: 60J05, 37A25.

\bigskip{}

\noindent\textbf{Keywords:} Nonlinear autoregressive model, autoregressive
conditional heteroskedasticity, ARCH, subgeometric ergodicity, polynomial
ergodicity, Markov chain, $\beta$-mixing.
\end{abstract}
\vfill{}

\pagebreak{}

\section{Introduction}

Let $X_{t}$ ($t=0,1,2,\ldots$) be a Markov chain on the state space
$\mathsf{X}$ and initialized from an $X_{0}$ following some initial
distribution. If the $n$-step probability measures $P^{n}(x\,;\,\cdot)=\Pr(X_{n}\in\cdot\mid X_{0}=x)$
converge in total variation norm $\lVert\,\cdot\,\rVert_{TV}$ to
the stationary probability measure $\pi$ at rate $r^{n}$ (for some
$r>1$), that is, 
\begin{equation}
\lim_{n\to\infty}r^{n}\lVert P^{n}(x\,;\,\cdot)-\pi(\cdot)\rVert_{TV}=0,\quad\textrm{ \ensuremath{\pi\:}a.e.,}\label{eq:Geom-erg}
\end{equation}
the Markov chain is said to be geometrically ergodic. When the convergence
in (\ref{eq:Geom-erg}) takes place at a suitably defined rate $r(n)$
slower than geometric, that is, 
\begin{equation}
\lim_{n\to\infty}r(n)\lVert P^{n}(x\,;\,\cdot)-\pi(\cdot)\rVert_{TV}=0,\quad\textrm{ \ensuremath{\pi\:}a.e.,}\label{eq:SubGeom-erg}
\end{equation}
the Markov chain is called subgeometrically ergodic. Examples of common
rates (where $c$ denotes a positive constant) include geometric (or
exponential) when $r(n)=e^{cn}=r^{n}$ ($r>1$), subexponential when
$r(n)=e^{cn^{\gamma}}$ ($0<\gamma<1$), polynomial when $r(n)=(1+n)^{c}$,
and logarithmic when $r(n)=(1+\ln(n))^{c}$. The authoritative and
classic reference to Markov chain theory is the monograph of \citet{meyn2009markov},
while an up-to-date treatment of subgeometric ergodicity can be found
in Chapters 16 and 17 of \citet*{douc2018markov}.

To give some background, the notion of subgeometric ergodicity was
introduced in the Markov chain literature in the 1980s when \citet{nummelin1983rate}
and \citet{tweedie1983criteria} obtained the first subgeometric ergodicity
results for general state space Markov chains. Subsequent work by
\citet{tuominen1994subgeometric}, \citet{fort2000vsubgeometric},
\citet{jarner2002polynomial}, \citet{fort2003polynomial}, and \citet*{douc2004practical}
lead to a formulation of a so-called drift condition to ensure subgeometric
ergodicity, paralleling the use of a Foster-Lyapunov drift condition
to establish geometric ergodicity (see, e.g., \citealp[Ch 15]{meyn2009markov}).
Various topics in probability theory and statistics have also been
considered under subgeometric assumptions; for instance, \citet*{douc2008bounds}
considered the central limit theorem and Berry-Esseen bounds, \citet{atchade2010limit}
the convergence of Markov chain Monte Carlo algorithms, \citet*{merlevede2011bernstein}
a Bernstein-type inequality, and \citet{meitz2021JAP} the rate of
$\beta$-mixing. In this paper we are interested in autoregressive
time series models. Results regarding the subgeometric ergodicity
of first-order autoregressions were obtained by \citet{tuominen1994subgeometric},
\citet{veretennikov2000polynomial}, \citet{fort2003polynomial},
\citet{douc2004practical}, \nocite{klokov2004sub,klokov2005subexponential}Klokov
and Veretennikov (2004, 2005), and \citet{klokov2007lower}, among
others, whereas results for more general higher-order autoregressions
were obtained by \citet{meitz2022subgear}. 

In this paper we consider subgeometric (specifically, polynomial)
ergodicity of autoregressive models with autoregressive conditional
heteroskedasticity (ARCH; \citealp{engle1982autoregressive}). The
previous works on subgeometrically ergodic autoregressions listed
above only considered the case of independent and identically distributed
(IID) errors, and allowing for conditionally heteroskedastic errors
considerably widens the scope of potential applications. This is particularly
important in applications using economic and financial time series
data. In the subgeometrically ergodic AR\textendash ARCH models we
consider, the conditional mean is similar to the (homoskedastic) AR
models already considered in \citet{meitz2022subgear}. The precise
model formulation will be given and motivated further in Section 2,
but we already note that the models we consider accommodate for behavior
similar to a unit root process for large values of the observed series
but almost no restrictions are placed on their dynamics for moderate
values of the observed series. The conditional variance is allowed
to follow a rather general nonlinear ARCH process. In our main result,
we show that the considered AR\textendash ARCH processes are, under
appropriate conditions, subgeometrically ergodic at a polynomial rate;\textcolor{red}{{}
}the convergence rate of $\beta$-mixing coefficients and finiteness
of certain moments are also obtained (for details, see Section 3.2).

The inclusion of ARCH (instead of IID) errors considerably complicates
the proofs of (sub)geometric ergodicity of nonlinear autoregressions.
Papers considering subgeometric ergodicity of homoskedastic autoregressions
were already listed above. Geometric ergodicity of nonlinear autoregressive
models with ARCH (or generalized ARCH) errors has previously been
considered by numerous authors; see, e.g., \citet{cline2004stability},
\nocite{meitz2008stability,meitz2010SPL} Meitz and Saikkonen (2008,
2010), and the many references therein. Compared to these two strands
of previous literature, the combination of the subgeometrically ergodic
type of nonlinear dynamics in the conditional mean with ARCH errors
leads to additional complications in the proofs. To appropriately
separate these two sources of dynamics we make use of a (relatively
unknown) extension of Bernoulli\textquoteright s inequality due to
\citet{FeffermanShapiro1972} (combined with Young's inequality),
and to control terms arising due to conditional heteroskedasticity
we devise a special matrix norm that is of a more complicated type
than the norms typically used when analysing the stability of nonlinear
time series models.

The rest of the paper is organized as follows. Section 2 introduces
the nonlinear AR\textendash ARCH model considered and states the assumptions
we employ. Results on subgeometric ergodicity are given in Section
3. In Section 4 we consider an empirical application of our model
to a daily time series of an energy sector volatility index. Section
5 concludes. All proofs are collected in an Appendix.

\section{Model}

\subsection{Conditional mean}

We consider the univariate process $y_{t}$ ($t=1,2,\ldots$) generated
by
\begin{equation}
y_{t}=\pi_{1}y_{t-1}+\cdots+\pi_{p-1}y_{t-p+1}+g(u_{t-1})+\sigma_{t}\varepsilon_{t},\label{NLAR}
\end{equation}
where $p\geq1$ is the autoregressive order, $u_{t}=y_{t}-\pi_{1}y_{t-1}-\cdots-\pi_{p-1}y_{t-p+1}$,
$g$ is a real-valued function, $\varepsilon_{t}$ is an IID error
term, and $\sigma_{t}=\sigma(\boldsymbol{y}_{t-1})$ is a positive
volatility term that depends on $p+q$ lagged values of $y_{t}$,
$\boldsymbol{y}_{t-1}=(y_{t-1},\ldots,y_{t-p-q})$, where $q\geq1$
is an ARCH order. For now, one concrete example of the volatility
term is a linear ARCH process, where $\sigma_{t}$ satisfies
\begin{equation}
\sigma_{t}^{2}=\omega+\alpha_{1}e_{t-1}^{2}+\cdots+\alpha_{q}e_{t-q}^{2}\label{ARCH(q)}
\end{equation}
and $e_{t}=y_{t}-\pi_{1}y_{t-1}-\cdots-\pi_{p-1}y_{t-p+1}-g(u_{t-1})$,
$\omega>0$, and $\alpha_{i}\geq0$ ($i=1,\ldots,q$); a more general
formulation for the conditional variance will be considered below.
Note that a compact expression for $e_{t}$ is $e_{t}=u_{t}-g(u_{t-1})$
so that equation (\ref{NLAR}) can be expressed as $u_{t}=g(u_{t-1})+\sigma_{t}\varepsilon_{t}$.
If $\pi_{1}=\cdots=\pi_{p-1}=0$ in equation (\ref{NLAR}), we have
$u_{t}=y_{t}$ so that the autoregressive order $p$ reduces to one
and equation (\ref{NLAR}) reduces to $y_{t}=g(y_{t-1})+\sigma_{t}\varepsilon_{t}$.

Our first assumption contains basic requirements for the error term
$\varepsilon_{t}$ and makes clear that the squared volatility, $\sigma_{t}^{2}$,
is the conditional variance of $y_{t}$ (when appropriate moments
exist). 
\begin{assumption}
\noindent $\{\varepsilon_{t},\,t=1,2,\ldots\}$ is a sequence of IID
random variables that is independent of $(y_{0},\ldots,y_{1-p-q})$,
has zero mean and unit variance, and the distribution of $\varepsilon_{1}$
has a (Lebesgue) density that is bounded away from zero on compact
subsets of $\mathbb{R}$.
\end{assumption}
\noindent Later on we introduce an assumption on the conditional variance
$\sigma_{t}^{2}$ which further restricts the moments of $\varepsilon_{t}$.

To further describe the conditional mean of the autoregressions we
consider, we next specify the conditions needed for the function $g$
in equation (\ref{NLAR}). The following assumption is a simplification
of Assumption 1 in \citet{meitz2022subgear} (the somewhat more general
formulation used therein is briefly discussed at the end of this subsection).
\begin{assumption}
\noindent \mbox{}

\noindent (i) The roots of the polynomial $\varpi(z)=1-\pi_{1}z-\cdots-\pi_{p-1}z^{p-1}$
lie outside the unit circle.

\noindent (ii) The function $g:\mathbb{R}\rightarrow\mathbb{R}$ in
(\ref{NLAR}) is measurable, locally bounded, and satisfies $\left|g(u)\right|\rightarrow\infty$
as $\left|u\right|\rightarrow\infty$, and there exist positive constants
$r$, $M_{0}$, $K_{0}$, and $0<\rho<2$ such that for all $u\in\mathbb{R}$
\begin{equation}
\left|g(u)\right|\leq\begin{cases}
(1-r\left|u\right|^{-\rho})\left|u\right| & \textrm{for }\left|u\right|\geq M_{0},\\
K_{0} & \textrm{for }\left|u\right|\leq M_{0}.
\end{cases}\label{Inequality_Ass 2}
\end{equation}
\end{assumption}
\noindent Assumption 2(i) corresponds to the conventional stationarity
condition of a linear autoregression in that it requires the roots
of the polynomial $\varpi(z)$ to lie outside the unit circle. In
the first-order case $p=1$, this condition becomes redundant because
then $\pi_{1}=\cdots=\pi_{p-1}=0$. Assumption 2(ii) is needed to
prove the subgeometric ergodicity of the process $y_{t}$, as already
done by \citet{fort2003polynomial} and \citet{douc2004practical}
in the first-order case $p=1$ and by \citet{meitz2022subgear} for
higher-order autoregressions.

We next provide some intuition and motivation for our model in (\ref{NLAR}).
To clarify the role of inequality (\ref{Inequality_Ass 2}) restricting
the function $g(\cdot)$, suppose Assumptions 1 and 2(i) hold but
instead of Assumption 2(ii) suppose the function $g(\cdot)$ were
linear with $g(u)=\pi_{0}u$ and $\pi_{0}\in[-1,1]$. Using the lag
operator $L$, equation (\ref{NLAR}) could then be written as
\begin{equation}
u_{t}-\pi_{0}u_{t-1}=(1-\pi_{0}L)(1-\pi_{1}L-\cdots-\pi_{p-1}L^{p-1})y_{t}=\sigma_{t}\varepsilon_{t},\label{LinearAR}
\end{equation}
that is, as the familiar linear AR($p$) model (with autoregressive
heteroskedasticity). Given Assumptions 1 and 2(i), the case $\pi_{0}\in(-1,1)$
corresponds to geometric ergodicity of $y_{t}$ and the cases $\pi_{0}=\pm1$
to non-ergodicity. Nonlinear functions $g(\cdot)$ satisfying Assumption
2(ii) provide a middle ground between these extreme cases of geometric
ergodicity and non-ergodicity. For instance, if $g(u)=(1-r\left|u\right|^{-\rho})u$
for $|u|>r^{1/\rho}$ and $g(u)=0$ otherwise ($r>0$, $0<\rho<2$),
then for any fixed $\pi_{0}\in(-1,1)$ and for all $u$ sufficiently
large in absolute value (i.e., for the values of $u$ that are crucial
for determining ergodicity),
\[
|\pi_{0}||u|<|g(u)|<|u|.
\]
The subgeometrically ergodic autoregressions we consider thus provide
one possibility for modeling small departures from unit root autoregressions.
Assumption 2(ii) implies that for large values of $|u_{t-1}|$, the
conditional mean of model (\ref{NLAR}) is close to that of an integrated
process (of order one). On the other hand, as inequality (\ref{Inequality_Ass 2})
restricts the function $g(\cdot)$ only for large values of its argument,
no restrictions (apart from the boundedness condition in (\ref{Inequality_Ass 2}))
are imposed when the argument takes values inside some bounded set
of values. Thus the autoregressions we consider may exhibit rather
arbitrary (stationary, unit root, explosive, nonlinear, etc.) behavior
for moderate values of the observed series. 

The autoregressions we consider are to some extent related to existing
models that have autoregressive roots near unity. To illustrate, when
$g(u)$ is as in the previous paragraph and we further set $p=r=\rho=1$,
the model in (\ref{NLAR}) simplifies to 
\[
y_{t}=\Bigl(1-\frac{1}{|y_{t-1}|}\Bigr)y_{t-1}+e_{t}\quad\text{when }|y_{t-1}|>1\quad\text{and}\quad y_{t}=e_{t}\quad\text{otherwise}
\]
where $e_{t}=\sigma_{t}\varepsilon_{t}$. In comparison, a prototypical
local-to-unity autoregression could be expressed as
\[
y_{t}=\Bigl(1-\frac{1}{T}\Bigr)y_{t-1}+e_{t},\quad t=1,\ldots,T,\quad\text{where \ensuremath{T} denotes the sample size}.
\]
Both of the above formulations involve an autoregressive coefficient
near unity, the former when the observed process takes on large (absolute)
values and the latter when the sample size is large. However, the
fact that the sample size is an essential part of local-to-unity autoregressions
makes them quite different from the autoregressions we consider \textemdash{}
in particular, the autoregressions we consider are ergodic. For more
details on local-to-unity autoregressions and other related models,
we refer the reader to the recent contributions of \citet{lieberman2020hybrid}
and \citet{phillips2023estimation} and the references therein.

Homoskedastic subgeometrically ergodic autoregressions satisfying
(a somewhat more general version of) Assumption 2 were already considered
by \citet{meitz2022subgear}. As many time series in economics, finance,
and other fields exhibit conditional heteroskedasticity, in this paper
we consider an extension to ARCH errors. In the homoskedastic case
considered in \citet{meitz2022subgear}, the term $g(u_{t-1})$ in
(\ref{NLAR}) was replaced with the more general formulation $u_{t-1}+\tilde{g}(y_{t-1},\ldots,y_{t-p})$
(with $\tilde{g}$ a real-valued function) to allow for more general
dependence on the past through the variables $y_{t-1},\ldots,y_{t-p}$
(and not only through the linear combination $u_{t-1}=y_{t-1}-\pi_{1}y_{t-2}-\cdots-\pi_{p-1}y_{t-p}$).
The present simpler formulation worked well in the empirical application
of Section 4 and in some other examples we tried out, and leads to
more transparent assumptions and streamlined proofs.

\subsection{Companion form}

To establish ergodicity, we need the companion form of the $(p+q)$-dimensional
process $\boldsymbol{y}_{t}=(\boldsymbol{y}_{1,t},\boldsymbol{y}_{2,t})$
with a $p$-dimensional $\boldsymbol{y}_{1,t}=(y_{t},\ldots,y_{t-p+1})$
and a $q$-dimensional $\boldsymbol{y}_{2,t}=(y_{t-p},\ldots,y_{t-p-q+1})$.
First we formulate the $p$-dimensional companion form related to
equation (\ref{NLAR}), which reads as
\[
\left[\begin{array}{c}
y_{t}\\
y_{t-1}\\
\vdots\\
\vdots\\
y_{t-p+1}
\end{array}\right]=\begin{bmatrix}\pi_{1} & \pi_{2} & \cdots & \pi_{p-1} & 0\\
1 & 0 & \cdots & 0 & 0\\
0 & \ddots & \ddots & \vdots & \vdots\\
\vdots & \ddots & \ddots & 0 & 0\\
0 & \cdots & 0 & 1 & 0
\end{bmatrix}\left[\begin{array}{c}
y_{t-1}\\
y_{t-2}\\
\vdots\\
\vdots\\
y_{t-p}
\end{array}\right]+g(u_{t-1})\left[\begin{array}{c}
1\\
0\\
\vdots\\
\vdots\\
0
\end{array}\right]+\sigma_{t}\varepsilon_{t}\left[\begin{array}{c}
1\\
0\\
\vdots\\
\vdots\\
0
\end{array}\right]
\]
or, denoting the matrix in this equation with $\boldsymbol{\Phi}$
and setting $\boldsymbol{\iota}_{p}=(1,0,\ldots,0)$ ($p\times1$),
as
\begin{equation}
\boldsymbol{y}_{1,t}=\boldsymbol{\Phi}\boldsymbol{y}_{1,t-1}+g(u_{t-1})\boldsymbol{\iota}_{p}+\sigma_{t}\varepsilon_{t}\boldsymbol{\iota}_{p}\label{Companion form 1}
\end{equation}
(when $p=1$, $\boldsymbol{\Phi}=0$ and $u_{t-1}=y_{t-1}$). As $\sigma_{t}=\sigma(\boldsymbol{y}_{t-1})$
depends on the whole $(p+q)$-dimensional vector $\boldsymbol{y}_{t-1}$,
we have to expand (\ref{Companion form 1}) to the $(p+q)$-dimensional
companion form
\begin{equation}
\left[\begin{array}{c}
\boldsymbol{y}_{1,t}\\
\boldsymbol{y}_{2,t}
\end{array}\right]=\begin{bmatrix}\begin{array}{cc}
\boldsymbol{\Phi} & \boldsymbol{0}_{p\times q}\end{array}\\
\begin{array}{ccc}
\hdashline\boldsymbol{0}_{q\times(p-1)} & I_{q} & \boldsymbol{0}_{q\times1}\end{array}
\end{bmatrix}\left[\begin{array}{c}
\boldsymbol{y}_{1,t-1}\\
\boldsymbol{y}_{2,t-1}
\end{array}\right]+g(u_{t-1})\boldsymbol{\iota}_{p+q}+\sigma_{t}\varepsilon_{t}\boldsymbol{\iota}_{p+q},\label{Companion form 2}
\end{equation}
where $I_{q}$ is the $(q\times q)$ identity matrix and $\boldsymbol{0}_{*\times*}$
denotes a matrix of zeros with the indicated dimensions (and $\boldsymbol{\iota}_{p+q}$
is defined in the obvious way). This shows that $\boldsymbol{y}_{t}$
is a Markov chain on $\mathbb{R}^{p+q}$.

In order to establish ergodicity we further transform the $p$-dimensional
companion form (\ref{Companion form 1}) in a way already used in
\citet[Sec 4]{meitz2022subgear}. To this end we define the matrices
\begin{equation}
\negthickspace\negthickspace\negthickspace\mathbf{A}=\setlength{\arraycolsep}{2pt}
\global\long\def\arraystretch{0.9}%
\begin{bmatrix}1 & -\pi_{1} & -\pi_{2} & \cdots & -\pi_{p-1}\\
0 & 1 & 0 & \cdots & 0\\
\vdots & \:\:\:\:\ddots & \ddots & \ddots & \vdots\\
0 & \:\:\: & \ddots & \ddots & 0\\
0 & \cdots & \cdots & 0 & 1
\end{bmatrix}\quad\text{and}\quad\mathbf{\Pi}=\mathbf{A}\boldsymbol{\Phi}\mathbf{A}^{-1}=\begin{bmatrix}0 & 0 & 0 & \cdots & 0\\
1 & \pi_{1} & \pi_{2} & \cdots & \pi_{p-1}\\
0 & 1 & 0 & \cdots & 0\\
\vdots & \ddots & \ddots & \ddots & \vdots\\
0 & \cdots & 0 & 1 & 0
\end{bmatrix}=\begin{bmatrix}0 & \boldsymbol{0}_{1\times(p-1)}\\
\boldsymbol{\iota}_{p-1} & \boldsymbol{\Pi}_{1}
\end{bmatrix},\negthickspace\negthickspace\negthickspace\negthickspace\label{Matrix Pi}
\end{equation}
\label{Cond exp V(y_1))}where $\mathbf{A}$ is nonsingular and $\boldsymbol{\Pi}_{1}$
is the $(p-1)\times(p-1)$ dimensional lower right hand corner of
$\mathbf{\Pi}$ (when $p=1$, $\mathbf{A}=1$ and $\mathbf{\Pi}=0$).
With these definitions equation (\ref{Companion form 1}) can be transformed
into
\begin{equation}
\mathbf{A}\boldsymbol{y}_{1,t}=\mathbf{\Pi}\mathbf{A}\boldsymbol{y}_{1,t-1}+g(u_{t-1})\boldsymbol{\iota}_{p}+\sigma_{t}\varepsilon_{t}\boldsymbol{\iota}_{p},\label{Companion form_A}
\end{equation}
where $\mathbf{A}\boldsymbol{y}_{1,t}=(u_{t},y_{t-1},\ldots,y_{t-p+1})$.
Now, for any $p$-dimensional vector $\boldsymbol{x}_{1}$, form the
partition $\boldsymbol{x}_{1}=(x_{1,1},\ldots,x_{1,p})=(x_{1,1},\boldsymbol{x}_{1,2})$
and define
\begin{equation}
\boldsymbol{z}(\boldsymbol{x}_{1})=\begin{bmatrix}z_{1}(\boldsymbol{x}_{1})\\
\boldsymbol{z}_{2}(\boldsymbol{x}_{1})
\end{bmatrix}=\mathbf{A}\boldsymbol{x}_{1}=\begin{bmatrix}x_{1,1}-\pi_{1}x_{1,2}-\cdots-\pi_{p-1}x_{1,p}\\
\boldsymbol{x}_{1,2}
\end{bmatrix}\label{z}
\end{equation}
(when $p=1$, $\boldsymbol{x}_{1,2}$ and $\boldsymbol{z}_{2}(\boldsymbol{x}_{1})$
are dropped). Using this notation equation (\ref{Companion form_A})
can be expressed as $\boldsymbol{z}(\boldsymbol{y}_{1,t})=\mathbf{\Pi}\boldsymbol{z}(\boldsymbol{y}_{1,t-1})+g(z_{1}(\boldsymbol{y}_{1,t-1}))\boldsymbol{\iota}_{p}+\sigma(\boldsymbol{y}_{t-1})\varepsilon_{t}\boldsymbol{\iota}_{p}$,
that is, as
\begin{align}
\begin{bmatrix}z_{1}(\boldsymbol{y}_{1,t})\\
\boldsymbol{z}_{2}(\boldsymbol{y}_{1,t})
\end{bmatrix} & =\setlength{\arraycolsep}{2pt}\begin{bmatrix}0 & \boldsymbol{0}_{1\times(p-1)}\\
\boldsymbol{\iota}_{p-1} & \boldsymbol{\Pi}_{1}
\end{bmatrix}\begin{bmatrix}z_{1}(\boldsymbol{y}_{1,t-1})\\
\boldsymbol{z}_{2}(\boldsymbol{y}_{1,t-1})
\end{bmatrix}+g(z_{1}(\boldsymbol{y}_{1,t-1}))\boldsymbol{\iota}_{p}+\sigma(\boldsymbol{y}_{t-1})\varepsilon_{t}\boldsymbol{\iota}_{p}\nonumber \\
 & =\begin{bmatrix}g(z_{1}(\boldsymbol{y}_{1,t-1}))+\sigma(\boldsymbol{y}_{t-1})\varepsilon_{t}\\
\boldsymbol{\Pi}_{1}\boldsymbol{z}_{2}(\boldsymbol{y}_{1,t-1})+z_{1}(\boldsymbol{y}_{1,t-1})\boldsymbol{\iota}_{p-1}
\end{bmatrix}.\label{Companion form_A2}
\end{align}
Here the first equation is in a form where the autoregressive order
is one and the volatility term is a function of the $(p+q)$-dimensional
vector $\boldsymbol{y}_{t-1}=(\boldsymbol{y}_{1,t-1},\boldsymbol{y}_{2,t-1})$
whereas the second equation involves the $p$-dimensional vector $\boldsymbol{y}_{1,t-1}$
only. 

By Assumption 2(i), the roots of the polynomial $\varpi(z)$ lie outside
the unit circle, so that the eigenvalues of the matrix $\boldsymbol{\Pi}_{1}$
in the second equation in (\ref{Companion form_A2}) are smaller than
one in absolute value. As is well known, this implies the existence
of a matrix norm of $\boldsymbol{\Pi}_{1}$ that is also smaller than
one. Specifically, for any vector norm $\Vert\cdot\Vert$, denote
by $\mnorm{\,\cdot\,}$ the corresponding induced matrix norm (\citealp[Defn 5.6.1]{HornJohnson2013});
that is, for any conformable square matrix $A$, set
\[
\mnorm{A}=\max_{\Vert\boldsymbol{x}\Vert=1}\Vert A\boldsymbol{x}\Vert.
\]
Then we obtain the following result (\citealp[Lemma 5.6.10]{HornJohnson2013}).
\begin{lem}
There exists a vector norm $\Vert\cdot\Vert_{*}$ and a corresponding
induced matrix norm $\mnorm{\,\cdot\,}_{*}$ such that $\mnorm{\boldsymbol{\Pi}_{1}}_{*}=\varpi<1$.
\end{lem}
The existence of an induced matrix norm with the property in the above
lemma is essential in our proofs. (When $p=1$, Assumption 2(i) and
Lemma 1 are redundant.) The norms $\Vert\cdot\Vert_{*}$ and $\mnorm{\,\cdot\,}_{*}$
are defined on $\mathbb{R}^{p-1}$ and $\mathbb{R}^{(p-1)\times(p-1)}$,
respectively, and they have been commonly used in time series models.
In the next subsection we introduce norms which are of a different
type. 

\subsection{Conditional variance}

The root condition of Assumption 2(i) and inequality (\ref{Inequality_Ass 2})
of Assumption 2(ii) are of major importance for establishing the stability
of our model. However, as these conditions only concern the conditional
mean, we need additional assumptions restricting the conditional variance
$\sigma_{t}^{2}$. As an extension of the basic ARCH model (\ref{ARCH(q)})
we consider a nonlinear formulation of the conditional variance defined
as 
\begin{equation}
\sigma_{t}^{2}=\zeta_{0,t-1}\omega+\alpha_{1}\zeta_{1,t-1}e_{t-1}^{2}+\cdots+\alpha_{q}\zeta_{q,t-1}e_{t-q}^{2},\label{NLARCH(q)}
\end{equation}
where $\zeta_{i,t-1}=\zeta_{i}(\boldsymbol{y}_{t-1})$ is a function
of $\boldsymbol{y}_{t-1}$ ($i=0,\ldots,q$) and otherwise the notation
is as in equation (\ref{ARCH(q)}) (including the conditions $\omega>0$
and $\alpha_{1},\ldots,\alpha_{q}\geq0$). When the functions $\zeta_{i,t-1}$
are the same for all $i=0,\ldots,q$ we remove the index $i$ and
use the notations $\zeta_{t-1}$ and $\zeta(\cdot)$. This is the
case in our empirical example where $\zeta_{t-1}=\zeta(y_{t-1})=1/(1+e^{-\gamma(y_{t-1}-a)})$
is a logistic function depending only on $y_{t-1}$. For possible
alternatives we consider a more general formulation and introduce
the following assumption.

\medskip{}

\begin{assumption}
\noindent In equation (\ref{NLARCH(q)}), the following conditions
are assumed. (i) The parameters $\omega,\alpha_{1},\ldots,\alpha_{q}$
satisfy $\omega>0$, $\alpha_{1},\ldots,\alpha_{q}\geq0$, and $\sum_{i=1}^{q}\alpha_{i}<1$.
(ii) For each $i=0,\ldots,q$, the function $\zeta_{i}$ takes values
in $(0,1]$.
\end{assumption}
\medskip{}

\noindent The above assumption includes the case $\zeta_{i}\equiv1$
for all $i$, which corresponds to the linear ARCH model (\ref{ARCH(q)}).
It covers also the above-mentioned logistic function.

Consider the $q$-dimensional process $\boldsymbol{\xi}_{t}=(e_{t}^{2},e_{t-1}^{2},\ldots,e_{t-q+1}^{2})$
($t\geq1$) with initial values $\boldsymbol{\xi}_{0}=(e_{0}^{2},\ldots,e_{-q+1}^{2})$
where $e_{0}^{2},\ldots,e_{-q+1}^{2}$ are functions of $\boldsymbol{y}_{0}$.
Inspired by \citet[Example 4.2]{cline2004stability} we now introduce
the following equation which is a straightforward implication of equation
(\ref{NLARCH(q)}) and the fact $\sigma_{t}\varepsilon_{t}=e_{t}$:
\[
\setlength{\arraycolsep}{3pt}\left[\begin{array}{c}
e_{t}^{2}\\
e_{t-1}^{2}\\
\vdots\\
\vdots\\
e_{t-q+1}^{2}
\end{array}\right]=\begin{bmatrix}\alpha_{1}\zeta_{1,t-1}\varepsilon_{t}^{2}\: & \alpha_{2}\zeta_{2,t-1}\varepsilon_{t}^{2} & \cdots & \alpha_{q-1}\zeta_{q-1,t-1}\varepsilon_{t}^{2}\: & \alpha_{q}\zeta_{q,t-1}\varepsilon_{t}^{2}\\
1 & 0 & \cdots & 0 & 0\\
0 & 1 & \ddots & \vdots & \vdots\\
\vdots & \ddots & \ddots & 0 & 0\\
0 & \cdots & 0 & 1 & 0
\end{bmatrix}\left[\begin{array}{c}
e_{t-1}^{2}\\
e_{t-2}^{2}\\
\vdots\\
\vdots\\
e_{t-q}^{2}
\end{array}\right]+\left[\begin{array}{c}
\zeta_{0,t-1}\varepsilon_{t}^{2}\omega\\
0\\
\vdots\\
\vdots\\
0
\end{array}\right]
\]
($t=1,2,\ldots$) or, more briefly, 
\begin{equation}
\boldsymbol{\xi}_{t}=\Lambda_{\zeta,t}\boldsymbol{\xi}_{t-1}+\boldsymbol{\omega}_{\zeta,t},\quad t=1,2,\ldots;\label{Companion form_A3}
\end{equation}

\noindent as $\boldsymbol{\xi}_{t}$ is a function of $\boldsymbol{y}_{t}$,
we occasionally write $\boldsymbol{\xi}_{t}=\boldsymbol{\xi}(\boldsymbol{y}_{t})$.
For later purposes we also note that due to the identities $\sigma_{t}\varepsilon_{t}=e_{t}$
and $e_{t}^{2}=\boldsymbol{\iota}'_{q}\boldsymbol{\xi}(\boldsymbol{y}_{t})$
we have 
\begin{equation}
\sigma_{t}^{2}=\sigma^{2}(\boldsymbol{y}_{t-1})=E[\boldsymbol{\iota}'_{q}\boldsymbol{\xi}(\boldsymbol{y}_{t})\mid\boldsymbol{y}_{t-1}].\label{Eq cond var}
\end{equation}
When there is need to make the dependence of $\Lambda_{\zeta,t}$
on $\boldsymbol{y}_{t-1}$ explicit we use the notation $\Lambda_{\zeta,t}(\boldsymbol{y}_{t-1})$
and replace the (random) argument $\boldsymbol{y}_{t-1}$ by a fixed
counterpart when needed. Specifically, $\Lambda_{\zeta,t}(\boldsymbol{x})$
means that the functions $\zeta_{i,t-1}=\zeta_{i,t-1}(\boldsymbol{y}_{t-1})$
used in $\Lambda_{\zeta,t}(\boldsymbol{y}_{t-1})$ are replaced by
$\zeta_{i,t}(\boldsymbol{x})$ for all $i=1,\ldots,q$, and the notations
$\sigma^{2}(\boldsymbol{x})$ and $\boldsymbol{\xi}(\boldsymbol{x})$
are used similarly.

We also define the matrices
\begin{equation}
\hspace{-5pt}\Lambda_{t}=\begin{bmatrix}\alpha_{1}\varepsilon_{t}^{2} & \alpha_{2}\varepsilon_{t}^{2} & \cdots & \alpha_{q-1}\varepsilon_{t}^{2} & \alpha_{q}\varepsilon_{t}^{2}\\
1 & 0 & \cdots & 0 & 0\\
0 & 1 & \ddots & \vdots & \vdots\\
\vdots & \ddots & \ddots & 0 & 0\\
0 & \cdots & 0 & 1 & 0
\end{bmatrix}\quad\text{and}\quad\Lambda=E[\Lambda_{t}]=\begin{bmatrix}\alpha_{1} & \alpha_{2} & \cdots & \alpha_{q-1} & \alpha_{q}\\
1 & 0 & \cdots & 0 & 0\\
0 & 1 & \ddots & \vdots & \vdots\\
\vdots & \ddots & \ddots & 0 & 0\\
0 & \cdots & 0 & 1 & 0
\end{bmatrix}.\label{Equations Lambda}
\end{equation}
Note that $\Lambda_{t}$ is obtained from the matrix $\Lambda_{\zeta,t}$
by choosing $\zeta_{i,t}=1$ for all $i=1,\ldots,q$. Similarly, we
denote $\boldsymbol{\omega}_{t}=(\omega_{t},0,\ldots,0)'$ with $\omega_{t}=\omega\varepsilon_{t}^{2}$
and $\boldsymbol{\omega}=E[\boldsymbol{\omega}_{t}]=(\omega,0,\ldots,0)'$.

In our proofs we need to appropriately control the size of the random
matrix $\Lambda_{t}$, and not just the size of the non-random matrix
$\Lambda=E[\Lambda_{t}]$. This is the reason why we next consider
vector and matrix norms more complicated than those in Lemma 1. To
this end, we first recall the definition of an $L$$^{p}$-norm (for
convenience, in this subsection only, we use the notation $p$ in
$L^{p}$-norms; elsewhere in the paper $p$ stands for the autoregressive
order in model (\ref{NLAR})). If $\Vert\cdot\Vert$ is any vector
norm on $\mathbb{R}^{q}$ and $\boldsymbol{v}$ is a $q$-dimensional
random vector, equation 
\[
\Vert\boldsymbol{v}\Vert_{L^{p}}=(E[\Vert\boldsymbol{v}\Vert^{p}])^{1/p}\qquad(1\leq p<\infty)
\]
defines an $L^{p}$-norm on the set of (equivalence classes of almost
surely equal) $q\times1$ random vectors that are $p$-integrable
(see, e.g., \citealp[Secs 5.1 and 5.2]{Dudley2004}). It may be worth
noting that for nonrandom vectors there is no difference between the
norms $\Vert\cdot\Vert$ and $\Vert\cdot\Vert_{L^{p}}$ but for random
vectors the outcome of $\Vert\cdot\Vert$ is random and that of $\Vert\cdot\Vert_{L^{p}}$
is nonrandom. This $L^{p}$-norm can be used to induce a norm for
random matrices; for the conventional non-random matrix case and for
the terminology used below, see \citet[Def. 5.6.1 and Sec 5.6.]{HornJohnson2013}.
Specifically, to define a generalized (non-submultiplicative) matrix
norm $\mnorm{\,\cdot\,}_{L^{p}}$, for any $q\times q$ random matrix
$A$ set
\begin{equation}
\mnorm{A}_{L^{p}}=\max_{\Vert\boldsymbol{x}\Vert_{L^{p}}=1}\Vert A\boldsymbol{x}\Vert_{L^{p}}=\max_{\Vert\boldsymbol{x}\Vert=1}\Vert A\boldsymbol{x}\Vert_{L^{p}}\qquad(\boldsymbol{x}\in\mathbb{R}^{q}),\label{Matrix norm_definition}
\end{equation}
where the latter equality holds as $\boldsymbol{x}$ is nonrandom.
This defines a generalized matrix norm\footnote{Axioms (1), (1a), (2), and (3) of a generalized matrix norm (see \citealp[pp. 340\textendash 341]{HornJohnson2013})
can be checked similarly as in the proof of Theorem 5.6.2(c) of the
same reference (replacing the norms $\Vert\cdot\Vert$ and $\mnorm{\,\cdot\,}$
therein with $\Vert\cdot\Vert_{L^{p}}$ and $\mnorm{\,\cdot\,}_{L^{p}}$,
replacing appropriate statements therein with their almost sure counterparts,
and using Minkowski's inequality as an additional justification for
axiom (3)).} on the set of (equivalence classes of almost surely equal) $q\times q$
random matrices with $p$-integrable entries; moreover, the norms
$\Vert\cdot\Vert$, $\Vert\cdot\Vert_{L^{p}}$, and $\mnorm{\,\cdot\,}_{L^{p}}$
are related by the inequality\footnote{Inequality (\ref{Ineq matrix norm}) can be verified analogously to
Theorem 5.6.2(b) of \citet{HornJohnson2013}.}
\begin{equation}
\Vert A\boldsymbol{x}\Vert_{L^{p}}\leq\mnorm{A}_{L^{p}}\Vert\boldsymbol{x}\Vert\qquad(\boldsymbol{x}\in\mathbb{R}^{q}).\label{Ineq matrix norm}
\end{equation}

We next state a high-level condition that assumes the existence of
a vector norm on $\mathbb{R}^{q}$ with particular additional properties.
(Primitive conditions ensuring this high-level assumption will be
given momentarily.) One of these properties is monotonicity in the
sense of Definition 5.4.18 of \citet{HornJohnson2013}: a vector norm
$\Vert\cdot\Vert$ is monotone if $\boldsymbol{x},\boldsymbol{y}\in\mathbb{R}^{q}$
satisfying $|x_{i}|\leq|y_{i}|$ for $i=1,\ldots,q$ always implies
that $\Vert\boldsymbol{x}\Vert\leq\Vert\boldsymbol{y}\Vert$. For
clarity, we use the notation $\Vert\cdot\Vert_{\bullet}$ for the
specific vector norm in the assumption below; similarly, we denote
the related $L^{p}$-norm by $\Vert\cdot\Vert_{\bullet L^{p}}$ and
the generalized matrix norm by $\mnorm{\,\cdot\,}_{\bullet L^{p}}$.
We also introduce two constants, $s_{0}\geq1$ and $b\geq1$, such
that 
\[
b=1\text{ when }s_{0}=1\quad\text{and}\quad b>(2s_{0}-\rho)/[s_{0}(2-\rho)]>1\text{ when }s_{0}>1
\]
(recall from Assumption 2 that $\rho\in(0,2)$ so that $2s_{0}>\rho$).
These constants are used in the next section where we establish our
ergodicity result and there the size of $s_{0}$ will have an effect
on the rate of convergence obtained and the order of moments that
are finite. The rather complex conditions required from the constant
$b$ are due to the connection between the conditional mean and ARCH
errors (this connection disappears when $s_{0}=1$ as it also does
in subgeometric homoskedastic autoregressions).
\begin{assumption}
\noindent Suppose there exists a vector norm $\Vert\cdot\Vert_{\bullet}$\textup{
on $\mathbb{R}^{q}$ that is (i) monotone and (ii) such that }$\mnorm{\Lambda_{t}}_{\bullet L^{bs_{0}}}=\lambda<1$,
where $b$ and $s_{0}$ are as described above.
\end{assumption}
\noindent This assumption tacitly requires that $E[|\varepsilon_{t}|^{2bs_{0}}]$
is finite, thereby strengthening Assumption 1 (when $s_{0}>1$). Assumption
4 is formulated in a way that is convenient in our proofs but is not
very transparent. The following lemma gives primitive conditions ensuring
that Assumption 4 holds (for a proof, see the appendix).
\begin{lem}
\noindent Suppose that Assumptions 1 and 3 hold and also that the
parameters $\alpha_{1},\ldots,\alpha_{q}$ in Assumption 3 satisfy
$\sum_{i=1}^{q}\alpha_{i}<1/\bar{\mu}_{2bs_{0}}$ where $\bar{\mu}_{2bs_{0}}=(E[|\varepsilon_{1}^{2}|^{bs_{0}}])^{1/bs_{0}}$.
Then Assumption 4 holds.
\end{lem}
To illustrate, consider the case $q=s_{0}=b=1$ so that the condition
in Lemma 2 reduces to the requirement $\alpha_{1}<1$. In geometrically
ergodic AR models with linear ARCH(1) errors, $\alpha_{1}<1$ is the
usual requirement for covariance stationarity while geometric ergodicity
can hold under even weaker conditions, such as $E[\ln(\alpha_{1}\varepsilon_{t}^{2})]<0$
(see, e.g., \citealp[Assumption 3 and Thm 1]{meitz2010SPL} for further
details). In the present setting the situation is different: as will
be seen in Section 3.2, condition $\alpha_{1}<1$ does not guarantee
a finite variance.

\section{Subgeometric ergodicity at a polynomial rate}

\subsection{Main result}

We now consider the stability of the model introduced in the previous
section. We begin with a brief account of some necessary Markov chain
concepts (for more comprehensive discussions, see \citet{meyn2009markov}
and \citet{douc2018markov} and also \citet[Sec 2]{meitz2022subgear}).
Let $X_{t}$ ($t=0,1,2,\ldots$) be a Markov chain on a general measurable
state space $(\mathsf{X},\mathcal{B}(\mathsf{X}))$ (with $\mathcal{B}(\mathsf{X})$
the Borel $\sigma$-algebra) and let $P^{n}(x\,;\,A)=\Pr(X_{n}\in A\mid X_{0}=x)$
signify its $n$-step transition probability measure. For an arbitrary
fixed measurable function $f:\mathsf{X}\rightarrow[1,\infty)$ and
for any signed measure $\mu$, define the $f$-norm $\left\Vert \mu\right\Vert _{f}$
as
\begin{equation}
\left\Vert \mu\right\Vert _{f}=\sup_{f_{0}:\left|f_{0}\right|\leq f}\left|\mu(f_{0})\right|,\label{eq:f-norm}
\end{equation}
where $\mu(f_{0})=\int_{x\in\mathsf{X}}f_{0}(x)\mu(dx)$ and the supremum
in (\ref{eq:f-norm}) runs over all measurable functions $f_{0}:\mathsf{X}\to\mathbb{R}$
such that $\left|f_{0}(x)\right|\leq f(x)$ for all $x\in\mathsf{X}$
(when $f\equiv1$, the $f$-norm $\left\Vert \mu\right\Vert _{f}$
reduces to the total variation norm $\left\Vert \mu\right\Vert _{TV}=\sup_{f_{0}:\left|f_{0}\right|\leq1}\left|\mu(f_{0})\right|$
used in (\ref{eq:Geom-erg}) and (\ref{eq:SubGeom-erg})). When the
$n$-step probability measures $P^{n}(x\,;\,\cdot)$ converge in $f$-norm
and at rate $r(n)$ to the stationary probability measure $\pi$ satisfying
$\pi(f)<\infty$, that is, \footnotetext[\numexpr\thefootnote+1\relax]{That is, the convergence in (\ref{f-ergodicity}) is required to hold for all $x\in\mathsf{X}$ except for those $x$ in a set that has probability zero with respect to the stationary measure $\pi$.}
\begin{equation}
\lim_{n\to\infty}r(n)\lVert P^{n}(x\,;\,\cdot)-\pi\rVert_{f}=0\qquad\text{for }\pi\text{-almost all }x\in\mathsf{X},~\footnotemark\label{f-ergodicity}
\end{equation}
we say that the Markov chain $X_{t}$ is ($f,r$)-ergodic; this implicitly
entails the existence of $\pi$ as well as certain moments as $\pi(f)<\infty$.
In the conventional geometrically ergodic case, $r(n)=r^{n}$ for
some $r>1$. To establish ($f,r$)-ergodicity, we use a so-called
drift condition defined as follows (here $\boldsymbol{1}_{S}(x)$
denotes the indicator function taking value one when $x$ belongs
to the set $S$ and zero elsewhere).\bigskip{}

\noindent \textbf{Condition D}. There exist a measurable function
$V\,:\,\mathsf{X}\rightarrow[1,\infty)$, a concave increasing continuously
differentiable function $\phi\,:\,[1,\infty)\rightarrow(0,\infty)$,
a measurable set $C$, and a finite constant $\tilde{b}$ such that
\begin{equation}
E\left[V(X_{1})\,\left|\,X_{0}=x\right.\right]\leq V(x)-\phi\left(V(x)\right)+\tilde{b}\boldsymbol{1}_{C}(x),\qquad x\in\mathsf{X}.\label{Drift condition}
\end{equation}

\bigskip{}

\noindent The idea is to verify this condition with suitable functions
$V$ and $\phi$, which together with some additional conditions ensures
the ($f,r$)-ergodicity of the process $X_{t}$; for more details,
see \citet[Thm 1]{meitz2022subgear}.

Now consider the stability of the Markov chain $\boldsymbol{y}_{t}$
on $\mathbb{R}^{p+q}$ given in (\ref{Companion form 2}). To define
the function $V$ in (\ref{Drift condition}), we use the functions
$z_{1}(\cdot)$, $\boldsymbol{z}_{2}(\cdot)$, and $\boldsymbol{\xi}(\cdot)$
in (\ref{z})\textendash (\ref{Companion form_A2}) and (\ref{Companion form_A3})
and the norms $\lVert\cdot\rVert_{*}$ and $\lVert\cdot\rVert_{\bullet}$
in Lemma 1 and Assumption 4. Set $\boldsymbol{x}=(x_{1},\ldots,x_{p+q})\in\mathbb{R}^{p+q}$
and decompose $\boldsymbol{x}$ to its $p$- and $q$-dimensional
components as  $\boldsymbol{x}=(\boldsymbol{x}_{1},\boldsymbol{x}_{2})$.
We define the function $V$ as
\begin{equation}
V(\boldsymbol{x})=1+|z_{1}(\boldsymbol{x}_{1})|^{2s_{0}}+s_{1}\lVert\boldsymbol{z}_{2}(\boldsymbol{x}_{1})\rVert_{*}^{2s_{0}\alpha}+s_{2}\lVert\boldsymbol{\xi}(\boldsymbol{x})\rVert_{\bullet}^{bs_{0}},\label{Eq V(x)}
\end{equation}
where $s_{0}$ and $b$ are defined above Assumption 4, $s_{1}$ and
$s_{2}$ are positive constants to be specified later (with $s_{1}$
small and $s_{2}$ large), and $\alpha=1-\rho/2s_{0}$ (recall from
Assumption 2 that $\rho\in(0,2)$ so that $\alpha\in(0,1)$). It may
be clarifying to note that when $p=1$, model (\ref{NLAR}) reduces
to $y_{t}=g(y_{t-1})+\sigma_{t}\varepsilon_{t}$; then we can set
$s_{1}=0$ and drop $\boldsymbol{z}_{2}(\boldsymbol{x}_{1})$ so that
the function $V$ in (\ref{Eq V(x)}) becomes $V(\boldsymbol{x})=1+|x_{1}|^{2s_{0}}+s_{2}\lVert\boldsymbol{\xi}(\boldsymbol{x})\rVert_{\bullet}^{bs_{0}}$.

To verify Condition D, we need to consider the conditional expectation
\begin{align}
E\left[V(\boldsymbol{y}_{1})\mid\boldsymbol{y}_{0}=\boldsymbol{x}\right]=1 & +E\left[\lvert z_{1}(\boldsymbol{y}_{1,1})\rvert^{2s_{0}}\mid\boldsymbol{y}_{0}=\boldsymbol{x}\right]+s_{1}E\left[\lVert\boldsymbol{z}_{2}(\boldsymbol{y}_{1,1})\rVert_{*}^{2s_{0}\alpha}\mid\boldsymbol{y}_{0}=\boldsymbol{x}\right]\nonumber \\
 & +s_{2}E\left[\lVert\boldsymbol{\xi}(\boldsymbol{y}_{1})\rVert_{\bullet}^{bs_{0}}\mid\boldsymbol{y}_{0}=\boldsymbol{x}\right],\label{Cond exp V(y_1)}
\end{align}
bound the conditional expectations on the right hand side of (\ref{Cond exp V(y_1)}),
and express these bounds in a way which conforms to inequality (\ref{Drift condition})
with the function $\phi$ satisfying the conditions required in Condition
D. These considerations, combined with the checking of some additional
technical conditions, lead to the following theorem (the proof can
be found in the Appendix). 
\begin{thm}
\noindent Consider the Markov chain $\boldsymbol{y}_{t}$ defined
in (\ref{Companion form 2}). Suppose that Assumptions 1\textendash 4
hold and that $V(\boldsymbol{x})$ is as in (\ref{Eq V(x)}). Then
$\boldsymbol{y}_{t}$ is ($f,r$)-ergodic with the polynomial convergence
rate $r(n)=n^{\delta-1}$ and the function $f$ given by $f(\boldsymbol{x})=V(\boldsymbol{x})^{1-\delta\rho/2s_{0}}$;
this result holds for any choice of $\delta\in[1,2s_{0}/\rho]$ and
for some (small enough) $s_{1}>0$ and some (large enough) $s_{2}>0$.
\end{thm}
Theorem 1 provides the first subgeometric ergodicity results for autoregressions
with autoregressive conditional heteroskedasticity. In this theorem,
the convergence rate $r(n)$ shows the speed at which the $n$-step
transition probability measures of the process $\boldsymbol{y}_{t}$
converge to the stationary probability measure. Due to the polynomial
convergence rate we therefore call the process $\boldsymbol{y}_{t}$
polynomially ergodic. Note also that the choice of $\delta$ in Theorem
1 allows for a trade-off between the rate of convergence and the size
of the $f$-norm.

\subsection{Discussion}

\paragraph{Geometric ergodicity.}

In previous literature, geometric ergodicity of nonlinear autoregressions
with ARCH errors has been considered using a variety of different
assumptions for the allowed nonlinear dynamics and for the required
moment conditions for the innovations; see, e.g., \citet{cline2004stability},
\citet{meitz2010SPL}, and the many references therein. 

\paragraph{Homoskedastic case.}

Theorem 1 remains valid also in the homoskedastic case (obtained by
setting $\alpha_{1}=\cdots=\alpha_{q}=0$). Previous polynomial ergodicity
results for homoskedastic autogressions were obtained by \citet[Sec 2.2]{fort2003polynomial}
and \citet[Thm 3]{meitz2022subgear}, and the above Theorem 1 provides
partial improvements over these earlier results in certain cases.
Assumptions and notation are slightly different in all the papers,
but (in the notation of the present paper) Theorem 1 improves earlier
results when $1\leq\rho<2$ and $1<s_{0}<2$. 

\paragraph{Proof strategy.}

The proof of Theorem 1 is also somewhat different from the previous
polynomial ergodicity results in \citet[Sec 2.2]{fort2003polynomial}
and \citet[Thm 3]{meitz2022subgear}. A rather obvious difference
is that these earlier results deal with homoskedastic autoregressions
whereas our model contains a nonlinear ARCH term, the size of which
is controlled with the special matrix norm defined in Assumption 4.
Regarding the conditional expectation, the mentioned earlier results
rely on Lemma 3 in \citet{fort2003polynomial} while our proof of
Theorem 1 avoids the use of this lemma, and instead makes use of a
(relatively unknown) extension of Bernoulli\textquoteright s inequality
due to \citet{FeffermanShapiro1972} (combined with Young's inequality).

\paragraph{Mixing and moment results. }

As already indicated in the Introduction, the polynomial ergodicity
result of Theorem 1 also implies that the process $\boldsymbol{y}_{t}$
is $\beta$-mixing (and hence $\alpha$-mixing). Moreover, the convergence
rate of the $\beta$-mixing coefficients $\beta(n)$ is given by the
fastest convergence rate, that is, $\lim_{n\rightarrow\infty}n^{2s_{0}/\rho-1}\beta(n)=0$.
For further details and justifications of these mixing results, see
\citet[Thm 2]{meitz2021JAP} and \citet[Sec 2]{meitz2022subgear}. 

Another consequence of Theorem 1 is that the stationary distribution
of $\boldsymbol{y}_{t}$ has finite moments up to order $2s_{0}-\rho$
(for a proof, see the Appendix). Note that depending on the values
of $s_{0}\geq1$ and $\rho\in(0,2)$, the order of these finite moments
may be very small; in particular, when $s_{0}=1$ we do not obtain
a finite variance.

\paragraph{Subexponential ergodicity.}

Theorem 1 concerns only polynomial ergodicity of subgeometric AR\textendash ARCH
models, and does not consider subexponential ergodicity (where the
rate $r(n)$ in (\ref{eq:SubGeom-erg}) equals, say, $e^{cn^{\gamma}}$
with $c>0$ and $0<\gamma<1$). The reason for this is that the properties
of ARCH-type models do not seem compatible with the moment requirements
needed for subexponential ergodicity. To elaborate on this, first
note that the previous results of \citet[Sec 3.3]{douc2004practical}
and \citet[Sec 4.1]{meitz2022subgear} on subexponential ergodicity
of homoskedastic nonlinear autoregressions (i) require the IID error
term to possess moments of \emph{all} orders and (ii) imply that the
observed process $y_{t}$ also has finite moments of \emph{all} orders.
(To provide some further details, (ii) is given as Corollary to Theorem
2 in \citet{meitz2022subgear}. As for (i), see Assumptions 3.3 and
2(a) of \citet{douc2004practical} and \citet{meitz2022subgear},
respectively. These assumptions require the IID error terms to be
sub-Weibull random variables, which in turn entails they possess moments
of all orders; see \citet[Defn 1 and Thm 1]{vladimirova2020sub} or
\citet[Defn 3 and Lemma 5]{wong2020lasso}.) 

The abovementioned moment requirements are in stark contrast to ARCH-type
models. For instance, in the simplest ARCH(1) model ($e_{t}=\sigma_{t}\varepsilon_{t}$,
$\sigma_{t}^{2}=\omega+\alpha_{1}e_{t-1}^{2}$, and $\varepsilon_{t}$
IID N(0,1)), the finiteness of moments of order $2r$ for $e_{t}$
($E[|e_{t}|^{2r}]<\infty$) is known to require the condition $\alpha_{1}^{r}E[|\varepsilon_{t}|^{2r}]<1$
(see, e.g., \citealp[Thm 2.1]{ling2002necessary} and \citealp[Example 6.1]{ling1999probabilistic}).
For integer values of $r$, this condition is equivalent with $\alpha_{1}<[(2r-1)!!]^{-1/r}=[1\cdot3\cdot\ldots\cdot(2r-1)]^{-1/r}$
and consequently \emph{all} moments of the ARCH process $e_{t}$ cannot
be finite unless $\alpha_{1}=0$. The situation is similar also in
more complicated (G)ARCH and AR\textendash (G)ARCH models (see, e.g.,
\citealp[Thm 2]{meitz2008ergodicity} and \citealp[Thm 1]{meitz2008stability},
respectively). This suggests that ARCH-type heteroskedastic errors
may not be compatible with the moment requirements needed for subexponential
ergodicity. 

\paragraph{Potential extensions.}

Extending our results to allow for GARCH (and not only ARCH) errors
would be interesting. However, previous literature suggests that studying
the stability of nonlinear AR\textendash GARCH models can be challenging.
Geometric ergodicity of nonlinear AR\textendash GARCH models has previously
been studied by \citet*{liu1997threshold}, \citet{ling1999probabilistic},
\citet{cline2007stability}, and \citet{meitz2008stability}; of these
articles, the former two are confined to threshold AR\textendash GARCH
models, whereas the latter two consider more general nonlinear autoregressions.
In the present setting, the autoregressive part of the model we consider
is rather general (the restrictions imposed on function $g(\cdot)$
in Assumption 2(ii) are quite mild, essentially restricting $g(\cdot)$
only for large values of its argument) and techniques used for threshold
models can not be applied. Using an approach similar to Cline's (2007)
appears challenging as the assumptions he employs are quite general
and appear difficult to verify (in fact, a threshold AR\textendash GARCH
model is the only example that is explicitly treated in his article).
On the other hand, \citet{meitz2008stability} require certain structure
and smoothness of the conditional mean (see Assumption 2 of their
paper) and it is not clear how to apply these results in the current
setting. As the extension to GARCH errors appears challenging, we
leave it for future research.

Another useful extension would be to consider the subgeometric ergodicity
of multivariate autoregressions with autoregressive conditional heteroskedasticity.
\citet[Sec 2.2]{fort2003polynomial} and \citet[Sec 3.3]{douc2004practical}
already studied multivariate first-order autoregressions with IID
errors and obtained results for polynomial and subexponential ergodicity,
respectively. In principle, generalizing these results to the higher-order
case with multivariate ARCH errors should be possible but it is not
immediate how to formulate a general model that would be both theoretically
manageable as well as useful in practical applications. We hope to
return to this issue in subsequent work.

\subsection{Examples}

The conditional mean of the model we have so far discussed is very
general, and we next consider some concrete illustrating examples.
The following two special cases were introduced in \citet[Sec 5]{meitz2022subgear}
in the case of a homoskedastic error term. We first consider a model
with a time-varying intercept term based on a logistic function and
specified as
\begin{equation}
y_{t}=\nu_{1}L(u_{t-1};\gamma,a_{1})+\nu_{2}(1-L(u_{t-1};\gamma,a_{2}))+y_{t-1}+\pi_{t-1}\Delta y_{t-1}+\cdots+\pi_{p-1}\Delta y_{t-p+1}+\sigma_{t}\varepsilon_{t},\label{LSTAR model}
\end{equation}
where $L(u;\gamma,a)=1/(1+e^{-\gamma(u-a)})$ is the logistic function
and the parameters $\gamma$, $a_{1}$, $a_{2}$ are assumed to satisfy
$\gamma>0$ and $a_{1}\leq a_{2}$, and $\nu_{1}$, $\nu_{2}$ are
assumed to satisfy $\nu_{1}<0<\nu_{2}$. Moreover, $\Delta$ signifies
the difference operator (so that $\Delta y_{t-1}=y_{t-1}-y_{t-2}$)
and the remaining notation is as in model (\ref{NLAR}). Arguments
similar to those in \citet[proof of Proposition 1]{meitz2022subgear}
can now be used to prove the following result (for details, see the
Appendix).
\begin{prop}
Consider the process $y_{t}$ defined in equation (\ref{LSTAR model})
and suppose that Assumptions 1, 2(i), 3, and 4 hold. Then, $\boldsymbol{y}_{t}$
is polynomially ergodic with convergence rate $r(n)=n^{2s_{0}-1}$
and finite moments up to order $2s_{0}-1$.
\end{prop}
The convergence rate presented in Proposition 1 also shows the rate
of $\beta$-mixing coefficients.

As another special case, we consider a model with a time-varying slope
term defined as
\begin{equation}
y_{t}=\pi_{t-1}y_{t-1}+\cdots+\pi_{p-1}y_{t-p+1}+S(u_{t-1})u_{t-1}+\sigma_{t}\varepsilon_{t},\label{ESTAR model}
\end{equation}
where $S(u_{t-1})$ is either $S_{1}(u_{t-1})=1-r_{0}/h(u_{t-1})$
or $S_{2}(u_{t-1})=\exp\{-r_{0}/h(u_{t-1})\}$ (with $r_{0}>0$) and
the function $h:\mathbb{R}\rightarrow(0,\infty)$ as defined in Proposition
2 of \citet[Sec 5.2]{meitz2022subgear}. In addition to a general
formulation of the function $h$ that proposition provides six special
cases of which two are $h(u)=1+|u-a|^{\rho}$ and $h(u)=(1+(u-a)^{2})^{\rho/2}$
(where $a\in\mathbb{R}$ and $\rho\in(0,2)$; see Assumption 2). Regarding
the remaining notation, it is as in model (\ref{NLAR}).

The following result can be established by using arguments similar
to those in the proof of Proposition 2 in \citet[Sec 5.2]{meitz2022subgear}
(for details, see the Appendix).
\begin{prop}
Consider the process $y_{t}$ defined in equation (\ref{ESTAR model})
and suppose that Assumptions 1, 2(i), 3, and 4 hold. Then, $\boldsymbol{y}_{t}$
is polynomially ergodic with convergence rate $r(n)=n^{2s_{0}/\rho-1}$
and finite moments up to order $2s_{0}-\rho$.
\end{prop}
The rate of $\beta$-mixing coefficients coincides with the rate given
in the proposition. As the function $h$ depends on the parameter
$\rho\in(0,2)$, the convergence rate in Proposition 2 differs from
that obtained in Proposition 1 except in the case $\rho=1$.

\section{Empirical application}

Although theoretical work on subgeometric ergodicity has been ongoing
for four decades, practical illustrations of (homoskedastic) subgeometrically
ergodic autoregressions have been scarce; we are not aware of \emph{any}
previous empirical applications of subgeometrically ergodic autoregressions
using real data. A small illustration of simulated data from one subgeometrically
ergodic autoregression is given in \citet[Sec 3]{fort2003polynomial}.
\citet[Sec 5]{meitz2022subgear} provide examples of some concrete
subgeometrically ergodic autoregressive time series models and illustrations
of a few simulated data series from them. These simulation exercises
suggest that subgeometrically ergodic autoregressions could be useful
when the observed time series bears some resemblance to unit root
type behavior and the autocorrelation function indicates very strong
persistence, but when the time series nevertheless exhibits eventual
mean-reverting behavior.\textcolor{red}{{} }The discussion in Section
2.1 around equation (\ref{LinearAR}) had a similar message, suggesting
these models could be seen as a middle ground between the extreme
cases of geometric ergodicity and non-ergodicity.

We next illustrate the use of subgeometrically ergodic AR\textendash ARCH
models in a small empirical example. Our aim is simply to provide
a proof of concept for the applicability of subgeometrically ergodic
AR\textendash ARCH models, illustrating that the model used fits the
data well. Further work is certainly needed to judge the usefulness
of these models in practical applications but we leave such more comprehensive
empirical applications for future research.

The data we employ consists of daily observations on the Chicago Board
Options Exchange energy sector volatility index (\url{fred.stlouisfed.org/series/VXXLECLS})
over the period 16 March 2011 through 31 December 2021 (a total of
2719 observations). This data series reflects energy sector risk and
is displayed in the top left graph of Figure 1 (the solid graph; the
dashed horizontal line shows the estimate $\hat{a}=25.366$, see (\ref{EmpAppModel})
and (\ref{EstimationResults}) below). The time series plot shows
signs of strong persistence, which is also reflected in the autocorrelation
function of the data shown in the top right graph of Figure 1.

\begin{figure}[tb]
\vspace*{-30pt}

\hfill{}%
\begin{minipage}[t][1\totalheight][c]{0.45\textwidth}%
\includegraphics[width=1.1\textwidth]{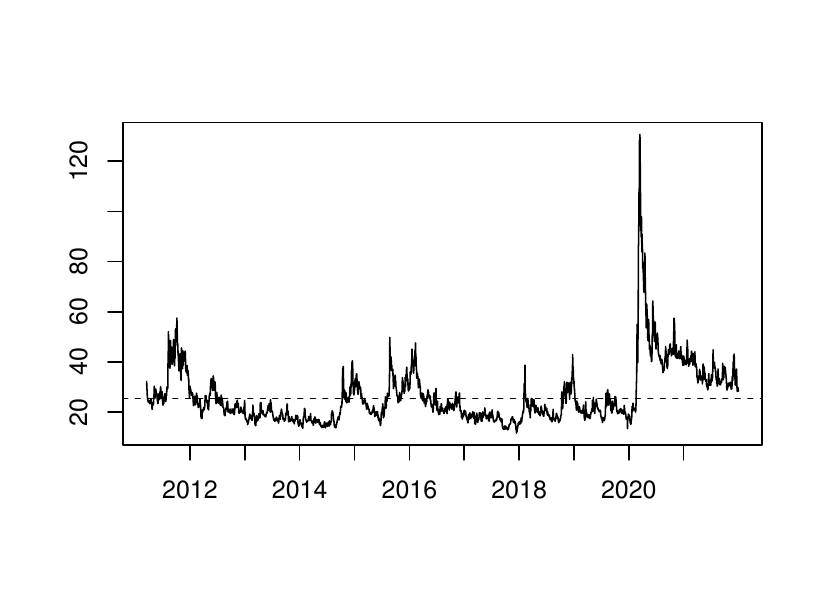}%
\end{minipage}\hfill{}%
\begin{minipage}[t][1\totalheight][c]{0.45\textwidth}%
\includegraphics[width=1.1\textwidth]{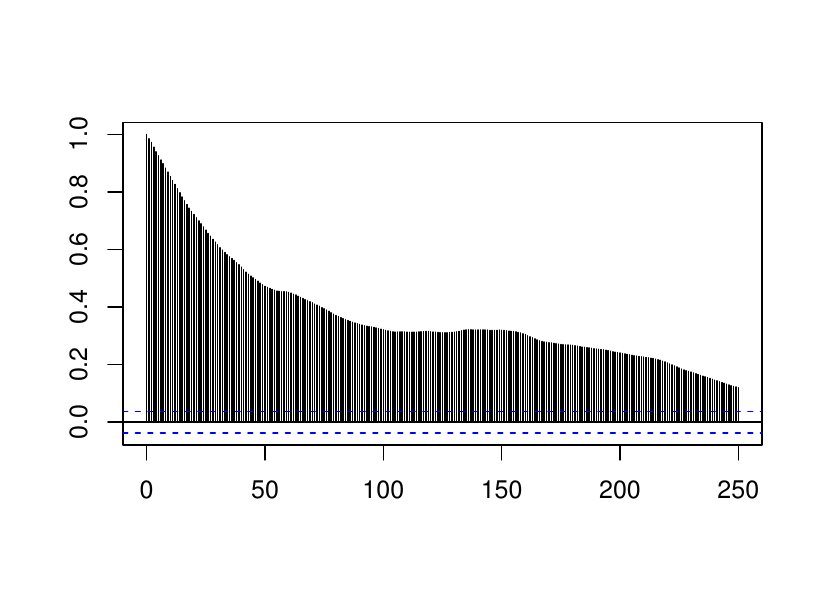}%
\end{minipage}\hfill{}\vspace*{-50pt}

\hfill{}%
\begin{minipage}[t][1\totalheight][c]{0.45\textwidth}%
\includegraphics[width=1.1\textwidth]{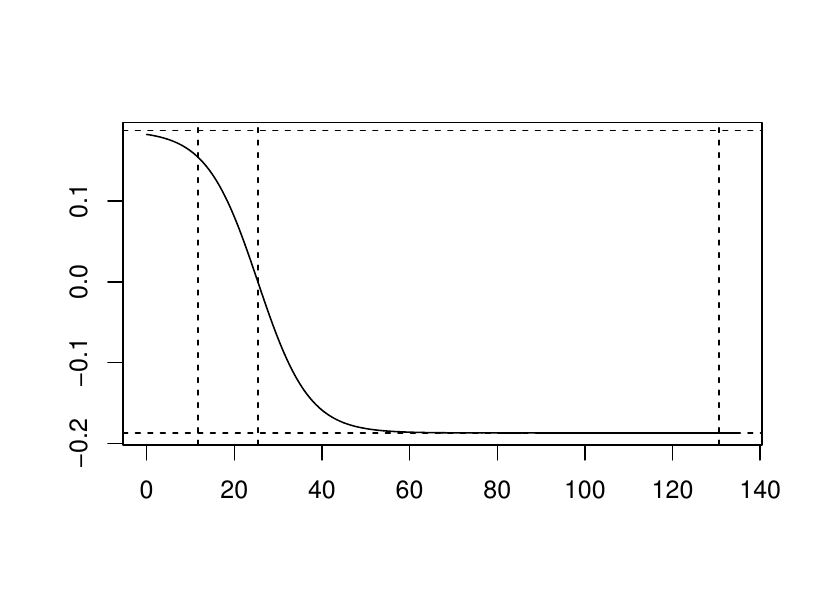}%
\end{minipage}\hfill{}%
\begin{minipage}[t][1\totalheight][c]{0.45\textwidth}%
\includegraphics[width=1.1\textwidth]{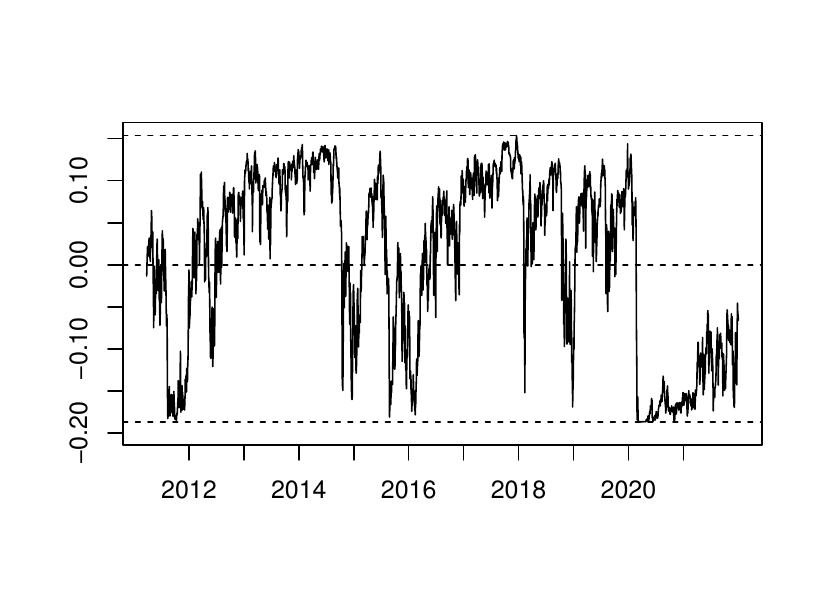}%
\end{minipage}\hfill{}\vspace*{-50pt}

\hfill{}%
\begin{minipage}[t][1\totalheight][c]{0.45\textwidth}%
\includegraphics[width=1.1\textwidth]{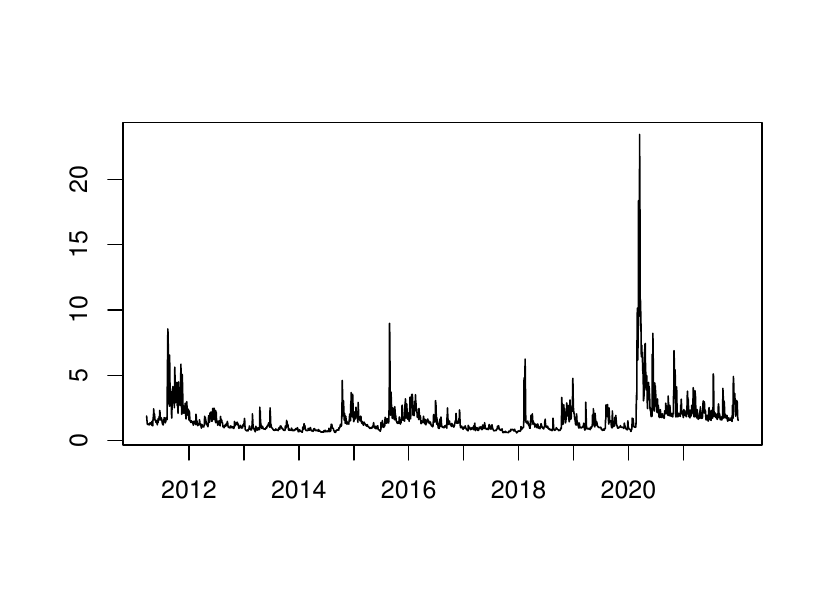}%
\end{minipage}\hfill{}%
\begin{minipage}[t][1\totalheight][c]{0.45\textwidth}%
\includegraphics[width=1.1\textwidth]{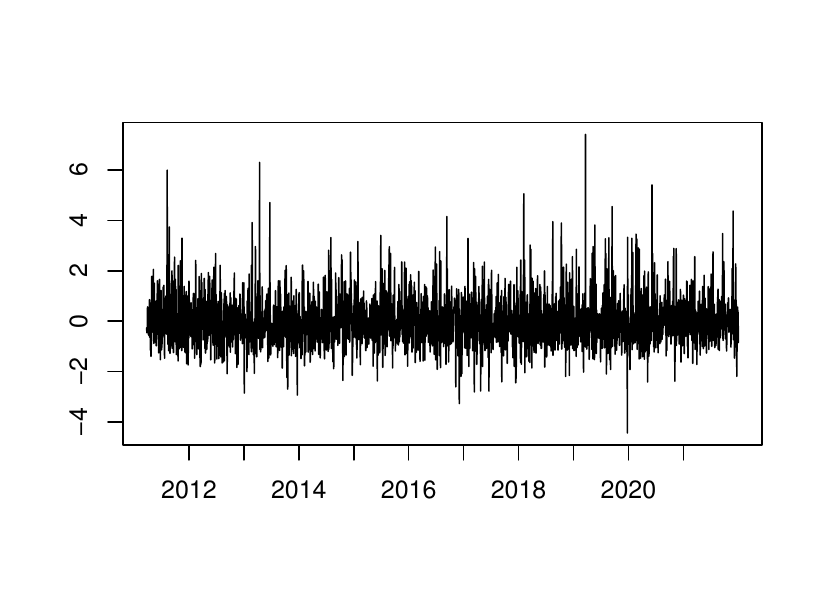}%
\end{minipage}\hfill{}\vspace*{-25pt}
\caption{\label{FigEx1} Top row: Daily observations on the Chicago Board Options
Exchange energy sector volatility index, 16 March 2011 \textendash{}
31 December 2021 (left); the corresponding autocorrelation function
(right). Middle row: The function $I(x)=-\nu L(x;\gamma,a)+\nu(1-L(x;\gamma,a))$
(left) and the corresponding time-varying intercept term $I(y_{t-1})$
(right), based on parameter estimates in (\ref{EstimationResults}).
Bottom row: The estimated volatility series $\hat{\sigma}_{t}$ (left)
and residual series $\hat{\varepsilon}_{t}$ (right), based on parameter
estimates in (\ref{EstimationResults}).}
\end{figure}
We model this data series using the parametric specification in (\ref{LSTAR model}).
As for the error distribution, after some experimentation a skew version
of the $t$-distribution due to \citet{jones2003skew} was found to
provide a good fit (in contrast, estimation with normal errors lead
to a distinct discrepancy between the residual distribution and the
Gaussian one). The density function of this distribution is
\[
f(x;c,d)=C_{c,d}^{-1}\biggl\{1+\frac{x}{(c+d+x^{\text{2}})^{1/2}}\biggr\}^{c+1/2}\biggl\{1-\frac{x}{(c+d+x^{\text{2}})^{1/2}}\biggr\}^{d+1/2},
\]
where $c$ and $d$ are positive parameters and $C_{c,d}=2^{c+d-1}B(c,d)(c+d)^{\text{1/2}}$
(with $B(\cdot,\cdot)$ denoting the beta function); the case $c=d$
results in a symmetric $t$-distribution with $2c$ degrees of freedom
and the cases $c<d$ and $c>d$ imply skewness to the left and right,
respectively. In our application, we use this distribution centralized
to have mean zero and standardized to have unit variance (i.e., in
the density function $x$ is replaced by $sx+m$ and $C_{c,d}^{-1}$
by $sC_{c,d}^{-1}$ where $m$ and $s^{2}$ denote the mean and variance,
see \citealp[Sec 2.1]{jones2003skew}; this requires that $c>1$ and
$d>1$, for a moment of order $k$ is finite when $c>k/2$ and $d>k/2$).

We estimate the model parameters using the method of maximum likelihood
and employ optimization routines in R. (We simply assume that standard
properties of maximum likelihood estimators hold and calculate standard
errors based on the standard formulas.) Trying out different model
orders lead to model (\ref{LSTAR model}) with order $p=1$ and with
a nonlinear ARCH term of order $q=3$ (with these choices the residual
diagnostics shown in Figure 2 in Appendix B indicated a very good
fit). Specifically, the considered model is
\begin{equation}
\begin{aligned}y_{t} & =y_{t-1}-\nu L(y_{t-1};\gamma,a)+\nu(1-L(y_{t-1};\gamma,a))+\sigma_{t}\varepsilon_{t}\\
\sigma_{t}^{2} & =(\omega+\alpha_{1}e_{t-1}^{2}+\alpha_{2}e_{t-2}^{2}+\alpha_{3}e_{t-3}^{2})L(y_{t-1};\gamma,a),
\end{aligned}
\label{EmpAppModel}
\end{equation}
where the errors $\varepsilon_{t}$ are IID$(0,1)$ and follow the
above described (centralized and standardized) skew $t$-distribution,
$L(y;\gamma,a)=1/(1+e^{-\gamma(y-a)})$ is the logistic function,
the parameters $\nu$ and $\gamma$ are positive, and $a\in\mathbb{R}$.
(We also tried a model where the logistic functions in the conditional
expectation and in the ARCH term were different but this extension
had only a minor effect on the results.) ML estimation (with the constraints
$\nu,\gamma,\omega>0$, $\alpha_{1},\alpha_{2},\alpha_{3}\geq0$,
and $\alpha_{1}+\alpha_{2}+\alpha_{3}<1$ that ensure polynomial ergodicity
by Proposition 1) leads to the following results:
\begin{equation}
\begin{aligned}y_{t} & =y_{t-1}-\underset{(0.040)}{0.187}L(y_{t-1};\underset{(0.018)}{0.171},\underset{(1.434)}{25.366})+\underset{(0.040)}{0.187}(1-L(y_{t-1};\underset{(0.018)}{0.171},\underset{(1.434)}{25.366}))+\hat{\sigma}_{t}\hat{\varepsilon}_{t}\\
\hat{\sigma}_{t}^{2} & =(\underset{(0.493)}{3.259}+\underset{(0.081)}{0.406}e_{t-1}^{2}+\underset{(0.066)}{0.310}e_{t-2}^{2}+\underset{(0.052)}{0.149}e_{t-3}^{2})L(y_{t-1};\underset{(0.018)}{0.171},\underset{(1.434)}{25.366}),
\end{aligned}
\label{EstimationResults}
\end{equation}
where the numbers in parenthesis are standard errors; estimates for
the parameters in the error distribution are $\hat{c}=3.551\:(0.422)$
and $\hat{d}=2.138\:(0.197)$.

To illustrate the conditional mean of the estimated model, consider
the function $I(x)=-\nu L(x;\gamma,a)+\nu(1-L(x;\gamma,a))$ and the
corresponding time-varying intercept term $I(y_{t-1})$ based on the
above parameter estimates. These are shown in the middle row of Figure
1. In the left panel, the two horizontal dashed lines show the minimum
and maximum $I(x)$ attains, while the three vertical dashed lines
indicate the minimum of the observed data series $y_{t}$ ($11.71$),
the estimate $\hat{a}=25.366$, and the maximum of $y_{t}$ ($130.61$).
On the right, the three horizontal dashed lines show the minimum and
maximum $I(y_{t-1})$ attains ($-0.187$ and $0.154$) and the origin.
Intuitively, when $y_{t-1}$ is close to $\hat{a}$ the time-varying
intercept term $I(y_{t-1})$ is close to zero and the conditional
mean of (\ref{EstimationResults}) corresponds to unit root type behavior
(without drift); when $y_{t-1}$ takes values clearly below/above
$\hat{a}$ the intercept $I(y_{t-1})$ is positive/negative and behavior
akin to a unit root process with increasing/decreasing drift occurs.

The left panel in the bottom row of Figure 1 displays the estimated
volatility series $\hat{\sigma}_{t}$. The variation of the volatility
over time is strong, and the large spikes in the volatility series
coincide with the large values in the observed series. The logistic
formulation of the conditional variance in (\ref{EstimationResults})
makes it possible for large observations to amplify volatility more
than a standard linear ARCH model would allow for. 

The right panel in the bottom row of Figure 1 shows the residual series
$\hat{\varepsilon}_{t}$. Four additional graphs analyzing the residuals
are available in Figure 2 in Appendix B: autocorrelation functions
of the residuals and of the squared residuals, together with a histogram
and a Q-Q plot. The autocorrelation functions reveal that the very
strong persistence present in the original series has been quite well
captured by the estimated subgeometrically ergodic AR\textendash ARCH
model (only three of the shown 100 autocorrelation coefficients are
barely outside the displayed critical values). The histogram and the
Q-Q plot indicate that the employed skew version of the $t$-distribution
fits well as only a few outlying observations deviate from the estimated
density function and the 45 degree line.

Note also that the estimated AR\textendash ARCH model satisfies the
requirements of a stationary and\textcolor{red}{{} }polynomially ergodic
process with finite absolute moments.\footnote{That is, the parameter estimates in (\ref{EstimationResults}) correspond
to a process satisfying the requirements of Proposition 1 with $s_{0}=1$.
(Note that these requirements are not satisfied with $s_{0}=1.5$
which would correspond to finite second moments of $y_{t}$.)} It may be interesting to note that estimation attempts using standard
linear ARMA(1,1)\textendash GARCH(1,1) models (with skew $t$ errors)
lead to estimated autoregressive coefficients in excess of $0.999$,
reflecting the very persistent nature of the data series apparent
from the time series and autocorrelation plots in the top row of Figure
1.

The primary purpose of this small empirical example was to demonstrate
what kind of time series could be modeled with subgeometrically ergodic
AR\textendash ARCH models. It is worth pointing out that such models
may work well even in cases where the graphs of the employed time
series and related autocorrelation functions look very different from
those displayed in Figure 1. 

\section{Conclusions}

In this paper, we examined the subgeometric ergodicity of nonlinear
autoregressive models with autoregressive conditional heteroskedasticity.
We provided conditions that ensured polynomial ergodicity of the considered
AR\textendash ARCH models. Our results generalized existing results
that assumed the error terms to be IID. The use of subgeometrically
ergodic AR\textendash ARCH models was illustrated in an empirical
example using energy sector volatility index data. 

Several future research topics could be entertained. In this paper
we have only considered ARCH-type conditional heteroskedasticity,
and extending the results to the generalized ARCH (GARCH) case would
be of interest. Subgeometric ergodicity of multivariate autoregressions
with autoregressive conditional heteroskedasticity is another interesting
topic left for future work. On the empirical side, further applied
work is certainly needed to judge the usefulness of subgeometrically
ergodic autoregressions (with or without ARCH) in practical applications.
For instance, providing more concrete advice on when to use subgeometrically
(rather than geometrically) ergodic autoregressions would be useful
for practitioners. Another question future applications should address
is whether subgeometrically ergodic autoregressions can outperform
relevant competing models in out-of-sample forecasting exercises.

\section*{Appendix A}

Appendix A contains the proofs of Lemma 2, Theorem 1, and Propositions
1 and 2 as well as details for the finiteness of moments in Section
3.2. 
\begin{proof}[\textbf{\emph{Proof of Lemma 2}}]
\noindent  Define the vector $(\bar{\alpha}_{1},\ldots,\bar{\alpha}_{q})=(\alpha_{1}\bar{\mu}_{2s_{0}b},\ldots,\alpha_{q}\bar{\mu}_{2s_{0}b})$
and let $\bar{\Lambda}$ denote the $q\times q$ matrix obtained by
replacing the first row of the matrix $\Lambda_{t}$ by $(\bar{\alpha}_{1},\ldots,\bar{\alpha}_{q})$.
By assumption, $\bar{\alpha}_{1},\ldots,\bar{\alpha}_{q}\geq0$ and
$\sum_{i=1}^{q}\bar{\alpha}_{i}<1$. These conditions ensure that
the polynomial $p(t)=t^{q}-\bar{\alpha}_{1}t^{q-1}-\cdots-\bar{\alpha}_{q}$
has all its roots inside the unit circle (if a root $t$ with $|t|\geq1$
existed, the contradiction $1=\bar{\alpha}_{1}/t+\cdots+\bar{\alpha}_{q}/t^{q}\leq\bar{\alpha}_{1}+\cdots+\bar{\alpha}_{q}$
would follow); this in turn implies that the matrix $\bar{\Lambda}$
has spectral radius $\rho(\bar{\Lambda})<1$ (see \citealp[pp. 194\textendash 195]{HornJohnson2013}).
Therefore the matrix $I_{q}-\bar{\Lambda}$ is invertible with $(I_{q}-\bar{\Lambda})^{-1}=\sum_{i=0}^{\infty}\bar{\Lambda}^{i}$.
Set $\boldsymbol{1}_{q}=(1,\ldots,1)$ ($q\times1$) and let $(\boldsymbol{x})^{abs}=(|x_{1}|,\ldots,|x_{q}|)$
($q\times1$) denote the elementwise absolute value of a vector $\boldsymbol{x}\in\mathbb{R}^{q}$.
We define the vector norm $\Vert\cdot\Vert_{\bullet}$ on $\mathbb{R}^{q}$
as $\Vert\boldsymbol{x}\Vert_{\bullet}=\boldsymbol{1}_{q}'(I_{q}-\bar{\Lambda})^{-1}(\boldsymbol{x})^{abs}$.
Note that as $(I_{q}-\bar{\Lambda})^{-1}=\sum_{i=0}^{\infty}\bar{\Lambda}^{i}$
with $\bar{\Lambda}$ having nonnegative (and also some strictly positive)
entries, $\Vert\boldsymbol{x}\Vert_{\bullet}=\boldsymbol{1}_{q}'(I_{q}-\bar{\Lambda})^{-1}(\boldsymbol{x})^{abs}>\boldsymbol{1}_{q}'(\boldsymbol{x})^{abs}=\Vert\boldsymbol{x}\Vert_{1}$
whenever $\boldsymbol{x}\neq0$ (here $\Vert\cdot\Vert_{1}$ denotes
the usual $l_{1}$ vector norm).

Now let $\boldsymbol{x}\neq0$ be arbitrary and consider $\Vert\Lambda_{t}\boldsymbol{x}\Vert_{\bullet}$.
To this end, note that $|\alpha_{1}\varepsilon_{t}^{2}x_{1}+\cdots+\alpha_{q}\varepsilon_{t}^{2}x_{q}|\leq\alpha_{1}\varepsilon_{t}^{2}|x_{1}|+\cdots+\alpha_{q}\varepsilon_{t}^{2}|x_{q}|$,
which implies that the elementwise inequality $(\Lambda_{t}\boldsymbol{x})^{abs}\leq\Lambda_{t}(\boldsymbol{x})^{abs}$
holds (with probability one; note that only the first elements differ).
Thus also
\[
\Vert\Lambda_{t}\boldsymbol{x}\Vert_{\bullet}=\boldsymbol{1}_{q}'(I_{q}-\bar{\Lambda})^{-1}(\Lambda_{t}\boldsymbol{x})^{abs}\leq\boldsymbol{1}_{q}'(I_{q}-\bar{\Lambda})^{-1}\Lambda_{t}(\boldsymbol{x})^{abs}.
\]
As $bs_{0}\geq1$, Minkowski\textquoteright s inequality and the definition
of the vector $(\bar{\alpha}_{1},\ldots,\bar{\alpha}_{q})$ yield
\[
E[\{\boldsymbol{1}_{q}'(I_{q}-\bar{\Lambda})^{-1}\Lambda_{t}(\boldsymbol{x})^{abs}\}^{bs_{0}}]^{1/bs_{0}}\leq\boldsymbol{1}_{q}'(I_{q}-\bar{\Lambda})^{-1}\bar{\Lambda}(\boldsymbol{x})^{abs},
\]
where, as $(I_{q}-\bar{\Lambda})^{-1}\bar{\Lambda}=(I_{q}-\bar{\Lambda})^{-1}-I_{q}$
and $\Vert\boldsymbol{x}\Vert_{\bullet}>\Vert\boldsymbol{x}\Vert_{1}$,
\[
\boldsymbol{1}_{q}'(I_{q}-\bar{\Lambda})^{-1}\bar{\Lambda}(\boldsymbol{x})^{abs}=\Vert\boldsymbol{x}\Vert_{\bullet}-\Vert\boldsymbol{x}\Vert_{1}=\Vert\boldsymbol{x}\Vert_{\bullet}(1-\Vert\boldsymbol{x}\Vert_{1}/\Vert\boldsymbol{x}\Vert_{\bullet})<\Vert\boldsymbol{x}\Vert_{\bullet}.
\]
These derivations establish that 
\[
\Vert\Lambda_{t}\boldsymbol{x}\Vert_{\bullet L^{bs_{0}}}=(E[\Vert\Lambda_{t}\boldsymbol{x}\Vert_{\bullet}^{bs_{0}}])^{1/bs_{0}}<\Vert\boldsymbol{x}\Vert_{\bullet}
\]
and that
\[
\mnorm{\Lambda_{t}}_{\bullet L^{bs_{0}}}=\max_{\Vert\boldsymbol{x}\Vert_{\bullet L^{bs_{0}}}=1}\Vert\Lambda_{t}\boldsymbol{x}\Vert_{\bullet L^{bs_{0}}}=\max_{\Vert\boldsymbol{x}\Vert_{\bullet}=1}\Vert\Lambda_{t}\boldsymbol{x}\Vert_{\bullet L^{bs_{0}}}<1.
\]
Finally, by its definition, it is clear that the vector norm $\Vert\cdot\Vert_{\bullet}$
is monotone.
\end{proof}
\bigskip{}

\noindent \textbf{Proof of Theorem 1: }For clarity, we break down
the long proof into several intermediate steps. 

\paragraph*{Step 1: Preliminaries.}

We first consider the function $V$ defined in (\ref{Eq V(x)}) and
the conditional expectation $E[V(\boldsymbol{y}_{1})\mid\boldsymbol{y}_{0}=\boldsymbol{x}]$.
As before, we decompose an $\boldsymbol{x}\in\mathbb{R}^{p+q}$ to
its $p$- and $q$-dimensional components as  $\boldsymbol{x}=(\boldsymbol{x}_{1},\boldsymbol{x}_{2})$;
similarly, we decompose $\boldsymbol{y}_{1}$ as $\boldsymbol{y}_{1}=(\boldsymbol{y}_{1,1},\boldsymbol{y}_{2,1})$.
For any $\boldsymbol{x}\in\mathbb{R}^{p+q}$, it is convenient to
define
\[
V_{1}(\boldsymbol{x}_{1})=\lvert z_{1}(\boldsymbol{x}_{1})\rvert^{2s_{0}},\qquad V_{2}(\boldsymbol{x}_{1})=s_{1}\lVert\boldsymbol{z}_{2}(\boldsymbol{x}_{1})\rVert_{*}^{2s_{0}\alpha},\qquad\text{and}\qquad V_{3}(\boldsymbol{x})=s_{2}\lVert\boldsymbol{\xi}(\boldsymbol{x})\rVert_{\bullet}^{bs_{0}}
\]
so that $V(\boldsymbol{x})=1+V_{1}(\boldsymbol{x}_{1})+V_{2}(\boldsymbol{x}_{1})+V_{3}(\boldsymbol{x})$
(when $p=1$, we can set $s_{1}=0$ and drop $\boldsymbol{z}_{2}(\boldsymbol{x}_{1})$
and $V_{2}$). We next consider the three conditional expectations
\begin{align}
E[V_{1}(\boldsymbol{y}_{1,1})\mid\boldsymbol{y}_{0}=\boldsymbol{x}] & =E[\lvert z_{1}(\boldsymbol{y}_{1,1})\rvert^{2s_{0}}\mid\boldsymbol{y}_{0}=\boldsymbol{x}]\label{Step1_EV1}\\
E[V_{2}(\boldsymbol{y}_{1,1})\mid\boldsymbol{y}_{0}=\boldsymbol{x}] & =E[s_{1}\lVert\boldsymbol{z}_{2}(\boldsymbol{y}_{1,1})\rVert_{*}^{2s_{0}\alpha}\mid\boldsymbol{y}_{0}=\boldsymbol{x}]\label{Step1_EV2}\\
E[V_{3}(\boldsymbol{y}_{1})\mid\boldsymbol{y}_{0}=\boldsymbol{x}] & =E[s_{2}\lVert\boldsymbol{\xi}(\boldsymbol{y}_{1})\rVert_{\bullet}^{bs_{0}}\mid\boldsymbol{y}_{0}=\boldsymbol{x}]\label{Step1_EV3}
\end{align}
related to functions $V_{1}$, $V_{2}$, and $V_{3}$. In Steps 2\textendash 4
below we establish that these conditional expectations can be bounded
from above using the following upper bounds
\begin{align}
E[\lvert z_{1}(\boldsymbol{y}_{1,1})\rvert^{2s_{0}}\mid\boldsymbol{y}_{0}=\boldsymbol{x}] & \leq|z_{1}(\boldsymbol{x}_{1})|^{2s_{0}}-\tilde{r}|z_{1}(\boldsymbol{x}_{1})|^{2s_{0}\alpha}+C\lVert\boldsymbol{\xi}(\boldsymbol{x})\rVert_{\bullet}^{bs_{0}}+C\label{Step1_EV1_Ineq}\\
E[s_{1}\lVert\boldsymbol{z}_{2}(\boldsymbol{y}_{1,1})\rVert_{*}^{2s_{0}\alpha}\mid\boldsymbol{y}_{0}=\boldsymbol{x}] & \leq s_{1}\left\Vert \boldsymbol{z}_{2}(\boldsymbol{x}_{1})\right\Vert _{*}^{2s_{0}\alpha}-\tilde{\varpi}s_{1}^{\alpha}\left\Vert \boldsymbol{z}_{2}(\boldsymbol{x}_{1})\right\Vert _{*}^{2s_{0}\alpha^{2}}+\tilde{s}_{1}\left|z_{1}(\boldsymbol{x}_{1})\right|^{2s_{0}\alpha}+C\label{Step1_EV2_Ineq}\\
E[s_{2}\lVert\boldsymbol{\xi}(\boldsymbol{y}_{1})\rVert_{\bullet}^{bs_{0}}\mid\boldsymbol{y}_{0}=\boldsymbol{x}] & \leq s_{2}\left\Vert \boldsymbol{\xi}(\boldsymbol{x})\right\Vert _{\bullet}^{bs_{0}}-\tilde{\lambda}s_{2}\left\Vert \boldsymbol{\xi}(\boldsymbol{x})\right\Vert _{\bullet}^{bs_{0}}+C,\label{Step1_EV3_Ineq}
\end{align}
where $\tilde{r},\tilde{\varpi},\tilde{s}_{1},\tilde{\lambda}>0$
with $\tilde{\lambda}<1$ and where $\tilde{s}_{1}$ can be made as
close to zero as desired by choosing a small enough $s_{1}$ (and
$\tilde{s}_{1}=0$ when $p=1$). Moreover, here and in what follows,
for simplicity we use $C$ to denote a finite positive constant whose
value may change from occurence to occurence (alternatively, we could
use $C_{1},C_{2},\ldots$). For brevity, we also often (but not always)
drop the argument $\boldsymbol{x}_{1}$ from $z_{1}(\boldsymbol{x}_{1})$
and $\boldsymbol{z}_{2}(\boldsymbol{x}_{1})$ and simply write $z_{1}$
and $\boldsymbol{z}_{2}$.

For ease of reference, we also note here that Assumption 4 allows
us to bound the conditional variance as follows. By the definition
of $\sigma_{t}^{2}$ in (\ref{NLARCH(q)}), Assumption 3, and definition
of $\boldsymbol{\xi}(\boldsymbol{y}_{t-1})$ in (\ref{Companion form_A3}),
$\sigma_{t}^{2}\leq\omega+e_{t-1}^{2}+\cdots+e_{t-q}^{2}=\omega+\Vert\boldsymbol{\xi}(\boldsymbol{y}_{t-1})\Vert_{1}$
(with $\Vert\cdot\Vert_{1}$ denoting the usual $l_{1}$ vector norm).
The equivalence of vector norms on $\mathbb{R}^{q}$ and the fact
that $\omega$ is a (finite) constant implies that (for some finite
constant $C$) 
\begin{equation}
\sigma_{t}^{2}=\sigma^{2}(\boldsymbol{y}_{t-1})\leq C(1+\Vert\boldsymbol{\xi}(\boldsymbol{y}_{t-1})\Vert_{\bullet})\,\,\text{a.s.}\quad\text{and}\quad\sigma^{2}(\boldsymbol{x})\leq C(1+\Vert\boldsymbol{\xi}(\boldsymbol{x})\Vert_{\bullet})\,\,\text{for all fixed \ensuremath{\boldsymbol{x}}}.\label{Ineq cond var}
\end{equation}

\paragraph*{Step 2: Upper bound for $V_{1}$.}

\noindent Using (\ref{Companion form_A2}) the conditional expectation
in (\ref{Step1_EV1}) can be expressed as
\[
E[V_{1}(\boldsymbol{y}_{1,1})\mid\boldsymbol{y}_{0}=\boldsymbol{x}]=E[|g(z_{1}(\boldsymbol{x}_{1}))+\sigma(\boldsymbol{x})\varepsilon_{1}|^{2s_{0}}].
\]
For any positive real number $Z,$ define the set $S_{1}(Z)=\{\boldsymbol{x}\in\mathbb{R}^{p+q}:|z_{1}(\boldsymbol{x}_{1})|\leq Z\}$
and let $S_{1}^{c}(Z)$ denote the complement of this set. 

First consider values of $\boldsymbol{x}$ such that $\boldsymbol{x}\in S_{1}^{c}(Z)$
so that $|z_{1}|=|z_{1}(\boldsymbol{x}_{1})|>Z$. Choose $Z$ large
enough to ensure that $g(z_{1})\neq0$ (Assumption 2) so that $|g(z_{1})+\sigma(\boldsymbol{x})\varepsilon_{1}|^{2s_{0}}$
can be written as 
\begin{equation}
|g(z_{1})|^{2s_{0}}\,|1+\sigma(\boldsymbol{x})\varepsilon_{1}/g(z_{1})|^{2s_{0}}.\label{Step2_1}
\end{equation}
We first bound the latter term in this expression using the following
extension of Bernoulli's inequality due to \citet{FeffermanShapiro1972}:
for any $a\geq2$, there exists positive numbers $A$ and $B$ such
that
\begin{equation}
|1+u|^{a}\leq1+au+Au^{2}+B|u|^{a}\label{FeffermanShapiroIneq}
\end{equation}
for all $u\in\mathbb{R}$. Using (\ref{FeffermanShapiroIneq}), the
latter term in (\ref{Step2_1}) is dominated by
\[
1+2s_{0}\frac{\sigma(\boldsymbol{x})}{g(z_{1})}\varepsilon_{1}+A\frac{\sigma(\boldsymbol{x})^{2}}{g(z_{1})^{2}}\varepsilon_{1}^{2}+B\frac{\sigma(\boldsymbol{x})^{2s_{0}}}{|g(z_{1})|^{2s_{0}}}|\varepsilon_{1}|^{2s_{0}}.
\]
This upper bound, (\ref{Step2_1}), the facts $E[\varepsilon_{1}]=0$
and $E[\varepsilon_{1}^{2}]=1$ (Assumption 1), and the notation $\mu_{2s_{0}}=E[|\varepsilon_{1}|^{2s_{0}}]$,
now yield
\[
E[V_{1}(\boldsymbol{y}_{1,1})\mid\boldsymbol{y}_{0}=\boldsymbol{x}]\leq|g(z_{1})|^{2s_{0}}+A|g(z_{1})|^{2s_{0}-2}\sigma(\boldsymbol{x})^{2}+B\sigma(\boldsymbol{x})^{2s_{0}}\mu_{2s_{0}}.
\]
Choose $Z$ large enough to ensure that $0<1-r|z_{1}|^{-\rho}<1$
and $|g(z_{1})|\leq(1-r|z_{1}|^{-\rho})|z_{1}|$ (Assumption 2). Using
the elementary inequalities $(1-u)^{a_{1}}\leq1-u$ and $(1-u)^{a_{2}}\leq1$
for all $0<u<1$, $a_{1}\geq1$, and $a_{2}\geq0$, and recalling
that $s_{0}\geq1$ and $\sigma^{2}(\boldsymbol{x})\leq C\left(1+\lVert\boldsymbol{\xi}(\boldsymbol{x})\rVert_{\bullet}\right)$
(see (\ref{Ineq cond var})), we obtain
\[
E[V_{1}(\boldsymbol{y}_{1,1})\mid\boldsymbol{y}_{0}=\boldsymbol{x}]\leq|z_{1}|^{2s_{0}}-r|z_{1}|^{2s_{0}-\rho}+C|z_{1}|^{2s_{0}-2}+C|z_{1}|^{2s_{0}-2}\lVert\boldsymbol{\xi}(\boldsymbol{x})\rVert_{\bullet}+C\lVert\boldsymbol{\xi}(\boldsymbol{x})\rVert_{\bullet}^{s_{0}}\mu_{2s_{0}}
\]
for some positive $C$ (by choosing $Z$ large enough, the constant
term on the dominant side has been absorbed into $|z_{1}|^{2s_{0}}$).
To merge the terms $-r|z_{1}|^{2s_{0}-\rho}$ and $C|z_{1}|^{2s_{0}-2}$,
by choosing $Z$ large enough to ensure that $C/r|z_{1}|^{2-\rho}<1$
we have 
\[
-r|z_{1}|^{2s_{0}-\rho}+C|z_{1}|^{2s_{0}-2}=-r|z_{1}|^{2s_{0}-\rho}(1-C/r|z_{1}|^{2-\rho})\leq-\hat{r}|z_{1}|^{2s_{0}-\rho}
\]
for some positive constant $\hat{r}$. Hence,
\begin{equation}
E[V_{1}(\boldsymbol{y}_{1,1})\mid\boldsymbol{y}_{0}=\boldsymbol{x}]\leq|z_{1}|^{2s_{0}}-\hat{r}|z_{1}|^{2s_{0}-\rho}+C|z_{1}|^{2s_{0}-2}\lVert\boldsymbol{\xi}(\boldsymbol{x})\rVert_{\bullet}+C\lVert\boldsymbol{\xi}(\boldsymbol{x})\rVert_{\bullet}^{s_{0}}\mu_{2s_{0}}\enskip\text{for all}\enskip\boldsymbol{x}\in S_{1}^{c}(Z).\label{Step2_2}
\end{equation}

Now consider values of $\boldsymbol{x}$ such that $\boldsymbol{x}\in S_{1}(Z)$.
As inequality (\ref{Inequality_Ass 2}) implies that $g(z_{1})$ is
bounded on $S_{1}(Z)$, triangle inequality and the elementary inequality
\begin{equation}
\biggl|\sum_{i=i}^{m}a_{i}\biggr|^{r}\leq c_{r}\sum_{i=i}^{m}\lvert a_{i}\rvert^{r}\quad\text{where }c_{r}=1\text{ for }0<r\leq1\text{ and }c_{r}=m^{r-1}\text{ for }r>1\label{c_r-ineq}
\end{equation}
for any real numbers $a_{1},\ldots,a_{m}$ (see, e.g., \citealp[p. 140]{Davidson1994})
imply that $|g(z_{1})+\sigma(\boldsymbol{x})\varepsilon_{1}|^{2s_{0}}$
is dominated by $C(1+\sigma^{2s_{0}}(\boldsymbol{x})|\varepsilon_{1}|^{2s_{0}})$
(for some $C>0$; we omit this statement from now on) for all $\boldsymbol{x}\in S_{1}(Z)$.
As $\mu_{2s_{0}}=E[|\varepsilon_{1}|^{2s_{0}}]$ and $\sigma^{2}(\boldsymbol{x})\leq C\left(1+\lVert\boldsymbol{\xi}(\boldsymbol{x})\rVert_{\bullet}\right)$,
it is seen that 
\begin{equation}
E[V_{1}(\boldsymbol{y}_{1,1})\mid\boldsymbol{y}_{0}=\boldsymbol{x}]\leq C(1+\lVert\boldsymbol{\xi}(\boldsymbol{x})\rVert_{\bullet}^{s_{0}}\mu_{2s_{0}})\quad\text{for all}\quad\boldsymbol{x}\in S_{1}(Z).\label{Step2_3}
\end{equation}

Combining (\ref{Step2_2}) and (\ref{Step2_3}), noting that $2s_{0}-\rho=2s_{0}\alpha$
(see the discussion following (\ref{Eq V(x)})), and merging constants,
we can conclude that for all $\boldsymbol{x}\in\mathbb{R}^{p+q}$,
\begin{equation}
E[V_{1}(\boldsymbol{y}_{1,1})\mid\boldsymbol{y}_{0}=\boldsymbol{x}]\leq|z_{1}|^{2s_{0}}-\hat{r}|z_{1}|^{2s_{0}\alpha}+C|z_{1}|^{2s_{0}-2}\lVert\boldsymbol{\xi}(\boldsymbol{x})\rVert_{\bullet}+C\lVert\boldsymbol{\xi}(\boldsymbol{x})\rVert_{\bullet}^{s_{0}}+C.\label{Step2_4}
\end{equation}
For future developments, it is convenient to further manipulate this
upper bound. First, consider the product $|z_{1}|^{2s_{0}-2}\Vert\boldsymbol{\xi}(\boldsymbol{x})\Vert_{\bullet}$
appearing in (\ref{Step2_4}) and momentarily focus on the case $s_{0}>1$
(when also $b>1$). Using Young's inequality (with exponents $bs_{0}/(bs_{0}-1)$
and $bs_{0}$) yields 
\[
|z_{1}|^{2s_{0}-2}\Vert\boldsymbol{\xi}(\boldsymbol{x})\Vert_{\bullet}\leq\frac{bs_{0}-1}{bs_{0}}|z_{1}|^{2s_{0}b(s_{0}-1)/(bs_{0}-1)}+\frac{1}{bs_{0}}\Vert\boldsymbol{\xi}(\boldsymbol{x})\Vert_{\bullet}^{bs_{0}}.
\]
Simple calculations show that the assumption $b>(2s_{0}-\rho)/[s_{0}(2-\rho)]$
implies that $2s_{0}b(s_{0}-1)/(bs_{0}-1)<2s_{0}\alpha$. Therefore
for some small positive $\varsigma$ 
\begin{equation}
|z_{1}|^{2s_{0}-2}\Vert\boldsymbol{\xi}(\boldsymbol{x})\Vert_{\bullet}\leq C(1+|z_{1}|^{2s_{0}\alpha-\varsigma}+\Vert\boldsymbol{\xi}(\boldsymbol{x})\Vert_{\bullet}^{bs_{0}});\label{Step2_5}
\end{equation}
clearly this upper bound also holds in the case $s_{0}=1$.

Second, consider the terms involving $|z_{1}|^{2s_{0}\alpha}$ in
(\ref{Step2_4}) and $|z_{1}|^{2s_{0}\alpha-\varsigma}$ in (\ref{Step2_5}).
By considering values of $|z_{1}|$ larger and smaller than some large
bound, it is straightforward to see that 
\[
C|z_{1}|^{2s_{0}\alpha-\varsigma}-\hat{r}|z_{1}|^{2s_{0}\alpha}=-\hat{r}(1-C/[\hat{r}|z_{1}|^{\varsigma}])|z_{1}|^{2s_{0}\alpha}\leq C-\tilde{r}|z_{1}|^{2s_{0}\alpha}
\]
for some positive constant $\tilde{r}$. Third, the term $\lVert\boldsymbol{\xi}(\boldsymbol{x})\rVert_{\bullet}^{s_{0}}$
appearing in (\ref{Step2_4}) is clearly dominated by a term of the
form $C+\lVert\boldsymbol{\xi}(\boldsymbol{x})\rVert_{\bullet}^{bs_{0}}$.

Inequality (\ref{Step2_4}) together with these additional manipulations
leads to the final upper bound 
\begin{equation}
E[V_{1}(\boldsymbol{y}_{1,1})\mid\boldsymbol{y}_{0}=\boldsymbol{x}]\leq|z_{1}|^{2s_{0}}-\tilde{r}|z_{1}|^{2s_{0}\alpha}+C\lVert\boldsymbol{\xi}(\boldsymbol{x})\rVert_{\bullet}^{bs_{0}}+C\label{Step2_Final}
\end{equation}
which holds for all $\boldsymbol{x}\in\mathbb{R}^{p+q}$.

\paragraph*{Step 3: Upper bound for $V_{2}$.}

\noindent Using (\ref{Companion form_A2}), we can express the conditional
expectation in (\ref{Step1_EV2}) as
\[
E[V_{2}(\boldsymbol{y}_{1,1})\mid\boldsymbol{y}_{0}=\boldsymbol{x}]=s_{1}\lVert\boldsymbol{\Pi}_{1}\boldsymbol{z}_{2}(\boldsymbol{x}_{1})+z_{1}(\boldsymbol{x}_{1})\boldsymbol{\iota}_{p-1}\rVert_{*}^{2s_{0}\alpha}.
\]
Recall that $\alpha=1-\rho/2s_{0}\in(0,1)$ (because $s_{0}\geq1$
and $\rho\in(0,2)$ by assumption) and that $\mnorm{\boldsymbol{\Pi}_{1}}_{*}\leq\varpi$
for some $0<\varpi<1$ (by Lemma 1). These facts together with elementary
inequalities (and dropping the argument $\boldsymbol{x}_{1}$ from
$z_{1}(\boldsymbol{x}_{1})$ and $\boldsymbol{z}_{2}(\boldsymbol{x}_{1})$)
imply that
\[
\left\Vert \boldsymbol{\Pi}_{1}\boldsymbol{z}_{2}+z_{1}\boldsymbol{\iota}_{p-1}\right\Vert _{*}^{\alpha}\leq\left\Vert \boldsymbol{\Pi}_{1}\boldsymbol{z}_{2}\right\Vert _{*}^{\alpha}+\left\Vert z_{1}\boldsymbol{\iota}_{p-1}\right\Vert _{*}^{\alpha}\leq\varpi^{\alpha}\left\Vert \boldsymbol{z}_{2}\right\Vert _{*}^{\alpha}+\left\Vert \boldsymbol{\iota}_{p-1}\right\Vert _{*}^{\alpha}\left|z_{1}\right|^{\alpha}.
\]
This, together with the convexity of the function $\left|x\right|\mapsto\left|x\right|^{2s_{0}}$
(recall that $s_{0}\geq1$ by assumption), imply that for any $\tau_{1}\in(0,1)$
and $\tau_{2}=1-\tau_{1}$, 
\begin{align}
s_{1}\left\Vert \boldsymbol{\Pi}_{1}\boldsymbol{z}_{2}+z_{1}\boldsymbol{\iota}_{p-1}\right\Vert _{*}^{2s_{0}\alpha} & \leq\left(\tau_{2}\frac{s_{1}^{1/2s_{0}}\varpi^{\alpha}}{\tau_{2}}\left\Vert \boldsymbol{z}_{2}\right\Vert _{*}^{\alpha}+\tau_{1}\frac{s_{1}^{1/2s_{0}}\left\Vert \boldsymbol{\iota}_{p-1}\right\Vert _{*}^{\alpha}}{\tau_{1}}\left|z_{1}\right|^{\alpha}\right)^{2s_{0}}\nonumber \\
 & \leq\tau_{2}\frac{s_{1}\varpi^{2s_{0}\alpha}}{\tau_{2}^{2s_{0}}}\left\Vert \boldsymbol{z}_{2}\right\Vert _{*}^{2s_{0}\alpha}+\tau_{1}\frac{s_{1}\left\Vert \boldsymbol{\iota}_{p-1}\right\Vert _{*}^{2s_{0}\alpha}}{\tau_{1}^{2s_{0}}}\left|z_{1}\right|^{2s_{0}\alpha}.\label{Step3_1}
\end{align}

Consider the former term on the dominant side of (\ref{Step3_1}).
Fix a $\tau_{2}$ such that $\tau_{2}\in(\varpi^{\alpha},1)$ and
set $\tilde{\varpi}=1-(\varpi^{\alpha}/\tau_{2})^{2s_{0}}\in(0,1)$.
Then the former term on the dominant side of (\ref{Step3_1}) satisfies
\begin{equation}
\tau_{2}\frac{s_{1}\varpi^{2s_{0}\alpha}}{\tau_{2}^{2s_{0}}}\left\Vert \boldsymbol{z}_{2}\right\Vert _{*}^{2s_{0}\alpha}=\tau_{2}s_{1}(1-\tilde{\varpi})\left\Vert \boldsymbol{z}_{2}\right\Vert _{*}^{2s_{0}\alpha}<s_{1}\left\Vert \boldsymbol{z}_{2}\right\Vert _{*}^{2s_{0}\alpha}-\tilde{\varpi}s_{1}\left\Vert \boldsymbol{z}_{2}\right\Vert _{*}^{2s_{0}\alpha}.\label{Step3_2}
\end{equation}
Suppose now that $s_{1}$ is any fixed (but potentially arbitrarily
small) positive number. If $\left\Vert \boldsymbol{z}_{2}\right\Vert _{*}$
is large enough to ensure that $s_{1}\left\Vert \boldsymbol{z}_{2}\right\Vert _{*}^{2s_{0}\alpha}\geq1$,
then $s_{1}\left\Vert \boldsymbol{z}_{2}\right\Vert _{*}^{2s_{0}\alpha}\geq s_{1}^{\alpha}\left\Vert \boldsymbol{z}_{2}\right\Vert _{*}^{2s_{0}\alpha^{2}}$
as $\alpha\in(0,1)$ and the right side of (\ref{Step3_2}) is dominated
by $s_{1}\left\Vert \boldsymbol{z}_{2}\right\Vert _{*}^{2s_{0}\alpha}-\tilde{\varpi}s_{1}^{\alpha}\left\Vert \boldsymbol{z}_{2}\right\Vert _{*}^{2s_{0}\alpha^{2}}$.
On the other hand, if $s_{1}\left\Vert \boldsymbol{z}_{2}\right\Vert _{*}^{2s_{0}\alpha}<1$
the right side of (\ref{Step3_2}) is bounded by a constant. Therefore
\[
\tau_{2}\frac{s_{1}\varpi^{2s_{0}\alpha}}{\tau_{2}^{2s_{0}}}\left\Vert \boldsymbol{z}_{2}\right\Vert _{*}^{2s_{0}\alpha}<s_{1}\left\Vert \boldsymbol{z}_{2}\right\Vert _{*}^{2s_{0}\alpha}-\tilde{\varpi}s_{1}^{\alpha}\left\Vert \boldsymbol{z}_{2}\right\Vert _{*}^{2s_{0}\alpha^{2}}+C.
\]

Now consider the latter term on the dominant side of (\ref{Step3_1}).
Choosing a small enough fixed $s_{1}$, this term can be made smaller
than $\tilde{s}_{1}\left|z_{1}\right|^{2s_{0}\alpha}$ where $\tilde{s}_{1}$
\textcolor{black}{can be chosen as close to zero as desired.} To summarize,
it holds that
\begin{align}
E[V_{2}(\boldsymbol{y}_{1,1})\mid\boldsymbol{y}_{0}=\boldsymbol{x}] & \leq s_{1}\left\Vert \boldsymbol{z}_{2}\right\Vert _{*}^{2s_{0}\alpha}-\tilde{\varpi}s_{1}^{\alpha}\left\Vert \boldsymbol{z}_{2}\right\Vert _{*}^{2s_{0}\alpha^{2}}+\tilde{s}_{1}\left|z_{1}\right|^{2s_{0}\alpha}+C\label{Step3_Final}
\end{align}
where $\tilde{\varpi}\in(0,1)$ and \textcolor{black}{the value of
}$\tilde{s}_{1}>0$\textcolor{black}{{} can be chosen as close to zero
as desired.}

\paragraph*{Step 4: Upper bound for $V_{3}$.}

By the definition of the function $\boldsymbol{\xi}$ in (\ref{Companion form_A3}),
$\boldsymbol{\xi}(\boldsymbol{y}_{1})=\Lambda_{\zeta,1}(\boldsymbol{y}_{0})\boldsymbol{\xi}(\boldsymbol{y}_{0})+\boldsymbol{\omega}_{\zeta,1}$.
We start by bounding both terms on the right hand side of this equality
and, for simplicity, remove the argument $\boldsymbol{y}_{0}$ and
instead use the notations $\Lambda_{\zeta,1}$ and $\boldsymbol{\xi}_{0}$.

First denote $v_{\zeta,1}=\varepsilon_{1}^{2}(\alpha_{1}\zeta_{1,0}e_{0}^{2}+\cdots+\alpha_{q}\zeta_{q,0}e_{1-q}^{2})$
and $v_{1}=\varepsilon_{1}^{2}(\alpha_{1}e_{0}^{2}+\cdots+\alpha_{q}e_{1-q}^{2})$.
Using the definitions of the matrices $\Lambda_{\zeta,1}$ and $\Lambda_{1}$
and the vector $\boldsymbol{\xi}_{0}$ (see (\ref{Companion form_A3})
and (\ref{Equations Lambda})) we then have
\[
\Lambda_{\zeta,1}\boldsymbol{\xi}_{0}=(v_{\zeta,1},e_{0}^{2},\ldots,e_{1-q}^{2})\quad\textrm{and}\quad\Lambda_{1}\boldsymbol{\xi}_{0}=(v_{1},e_{0}^{2},\ldots,e_{1-q}^{2}),
\]
where all components of both vectors are nonnegative. As $\zeta_{i,0}\in(0,1]$
for all $i=1,\ldots,q$ by assumption, we have $v_{\zeta,1}\leq v_{1}$
(a.s.). The monotonicity of the norm $\Vert\cdot\Vert_{\bullet}$
required in Assumption 4 now implies that $\left\Vert \Lambda_{\zeta,1}\boldsymbol{\xi}_{0}\right\Vert _{\bullet}\leq\left\Vert \Lambda_{1}\boldsymbol{\xi}_{0}\right\Vert _{\bullet}$
(a.s.) (see the discussion preceding Assumption 4). Regarding the
vector $\boldsymbol{\omega}_{\zeta,1}$, its first component is $\zeta_{0,0}(\boldsymbol{y}_{0})\varepsilon_{1}^{2}\omega\leq\varepsilon_{1}^{2}\omega$
(a.s.) and the other components are zero, so that the monotonicity
of the norm $\Vert\cdot\Vert_{\bullet}$ shows that $\lVert\boldsymbol{\omega}_{\zeta,1}\rVert_{\bullet}\leq\lVert\boldsymbol{\omega}_{1}\rVert_{\bullet}=\varepsilon_{1}^{2}\lVert\boldsymbol{\omega}\rVert_{\bullet}$
(a.s.) where $\boldsymbol{\omega}=(\omega,0,\ldots,0)$.

The preceding discussion together with the triangle inequality now
yields, with probability one,
\[
\Vert\boldsymbol{\xi}(\boldsymbol{y}_{1})\Vert_{\bullet}=\lVert\Lambda_{\zeta,1}(\boldsymbol{y}_{0})\boldsymbol{\xi}(\boldsymbol{y}_{0})+\boldsymbol{\omega}_{\zeta,1}\rVert_{\bullet}\leq\lVert\Lambda_{\zeta,1}(\boldsymbol{y}_{0})\boldsymbol{\xi}(\boldsymbol{y}_{0})\rVert_{\bullet}+\lVert\boldsymbol{\omega}_{\zeta,1}\rVert_{\bullet}\leq\lVert\Lambda_{1}\boldsymbol{\xi}(\boldsymbol{y}_{0})\rVert_{\bullet}+\varepsilon_{1}^{2}\lVert\boldsymbol{\omega}\rVert_{\bullet}.
\]
Using the notation $\bar{\mu}_{2bs_{0}}=(E[|\varepsilon_{0}|^{2bs_{0}}])^{1/bs_{0}}$
and Minkowski\textquoteright s inequality we find that
\[
\left(E[\lVert\boldsymbol{\xi}(\boldsymbol{y}_{1})\rVert_{\bullet}^{bs_{0}}\mid\boldsymbol{y}_{0}=\boldsymbol{x}]\right)^{1/bs_{0}}\leq\left(E[\lVert\Lambda_{1}\boldsymbol{\xi}(\boldsymbol{x})\rVert_{\bullet}^{bs_{0}}]\right)^{1/bs_{0}}+\bar{\mu}_{2bs_{0}}\lVert\boldsymbol{\omega}\rVert_{\bullet}.
\]
By inequality (\ref{Ineq matrix norm}) and Assumption 4, the first
term on the dominant side satisfies
\[
\left(E[\lVert\Lambda_{1}\boldsymbol{\xi}(\boldsymbol{x})\rVert_{\bullet}^{bs_{0}}]\right)^{1/bs_{0}}=\lVert\Lambda_{1}\boldsymbol{\xi}(\boldsymbol{x})\rVert_{\bullet L^{bs_{0}}}\leq\mnorm{\Lambda_{1}}_{\bullet L^{bs_{0}}}\lVert\boldsymbol{\xi}(\boldsymbol{x})\rVert_{\bullet}=\lambda\lVert\boldsymbol{\xi}(\boldsymbol{x})\rVert_{\bullet}
\]
with $\lambda<1$. The preceding steps imply that
\[
\left(E[\lVert\boldsymbol{\xi}(\boldsymbol{y}_{1})\rVert_{\bullet}^{bs_{0}}\mid\boldsymbol{y}_{0}=\boldsymbol{x}]\right)^{1/bs_{0}}\leq\lambda\lVert\boldsymbol{\xi}(\boldsymbol{x})\rVert_{\bullet}+\bar{\mu}_{2bs_{0}}\lVert\boldsymbol{\omega}\rVert_{\bullet}=\lambda\lVert\boldsymbol{\xi}(\boldsymbol{x})\rVert_{\bullet}\left(1+\frac{\bar{\mu}_{2bs_{0}}\lVert\boldsymbol{\omega}\rVert_{\bullet}}{\lambda\lVert\boldsymbol{\xi}(\boldsymbol{x})\rVert_{\bullet}}\right)
\]
and, using (\ref{Step1_EV3}) and the notation $\hat{\lambda}=\lambda^{bs_{0}}<1$,
we obtain
\begin{align*}
E[V_{3}(\boldsymbol{y}_{1}) & \mid\boldsymbol{y}_{0}=\boldsymbol{x}]\leq s_{2}\hat{\lambda}\lVert\boldsymbol{\xi}(\boldsymbol{x})\rVert_{\bullet}^{bs_{0}}\left(1+\frac{\bar{\mu}_{2bs_{0}}\lVert\boldsymbol{\omega}\rVert_{\bullet}}{\lambda\lVert\boldsymbol{\xi}(\boldsymbol{x})\rVert_{\bullet}}\right)^{bs_{0}}\\
 & \qquad\qquad\;=s_{2}\lVert\boldsymbol{\xi}(\boldsymbol{x})\rVert_{\bullet}^{bs_{0}}-\left\{ 1-\hat{\lambda}\left(1+\frac{\bar{\mu}_{2bs_{0}}\lVert\boldsymbol{\omega}\rVert_{\bullet}}{\lambda\lVert\boldsymbol{\xi}(\boldsymbol{x})\rVert_{\bullet}}\right)^{bs_{0}}\right\} s_{2}\left\Vert \boldsymbol{\xi}(\boldsymbol{x})\right\Vert _{\bullet}^{bs_{0}}.
\end{align*}

\noindent As $\hat{\lambda}\in(0,1)$, we can choose a $\xi>0$ such
that the term in curly brackets is larger than some $\tilde{\lambda}\in(0,1)$
whenever $\left\Vert \boldsymbol{\xi}(\boldsymbol{x})\right\Vert _{\bullet}>\xi$.
Therefore, whenever $\left\Vert \boldsymbol{\xi}(\boldsymbol{x})\right\Vert _{\bullet}>\xi$,
we have
\[
E[V_{3}(\boldsymbol{y}_{1})\mid\boldsymbol{y}_{0}=\boldsymbol{x}]\leq s_{2}\left\Vert \boldsymbol{\xi}(\boldsymbol{x})\right\Vert _{\bullet}^{bs_{0}}-\tilde{\lambda}s_{2}\left\Vert \boldsymbol{\xi}(\boldsymbol{x})\right\Vert _{\bullet}^{bs_{0}}.
\]
On the other hand, whenever $\lVert\boldsymbol{\xi}(\boldsymbol{x})\rVert_{\bullet}\leq\xi$
the previous derivations also make it clear that $E[V_{3}(\boldsymbol{y}_{1})\mid\boldsymbol{y}_{0}=\boldsymbol{x}]$
is bounded by some constant. Therefore for all $\boldsymbol{x}\in\mathbb{R}^{p+q}$,
\begin{equation}
E[V_{3}(\boldsymbol{y}_{1})\mid\boldsymbol{y}_{0}=\boldsymbol{x}]\leq s_{2}\left\Vert \boldsymbol{\xi}(\boldsymbol{x})\right\Vert _{\bullet}^{bs_{0}}-\tilde{\lambda}s_{2}\left\Vert \boldsymbol{\xi}(\boldsymbol{x})\right\Vert _{\bullet}^{bs_{0}}+C.\label{Step4_Final}
\end{equation}

\paragraph*{Step 5: Upper bound for $V$.}

\noindent We next combine the upper bounds (\ref{Step2_Final}), (\ref{Step3_Final}),
and (\ref{Step4_Final}) derived in Steps 2\textendash 4 to obtain
\begin{align}
E[V(\boldsymbol{y}_{1})\mid\boldsymbol{y}_{0}=\boldsymbol{x}] & \leq1+|z_{1}|^{2s_{0}}-\tilde{r}|z_{1}|^{2s_{0}\alpha}+C\lVert\boldsymbol{\xi}(\boldsymbol{x})\rVert_{\bullet}^{bs_{0}}\nonumber \\
 & \qquad+s_{1}\left\Vert \boldsymbol{z}_{2}\right\Vert _{*}^{2s_{0}\alpha}-\tilde{\varpi}s_{1}^{\alpha}\left\Vert \boldsymbol{z}_{2}\right\Vert _{*}^{2s_{0}\alpha^{2}}+\tilde{s}_{1}\left|z_{1}\right|^{2s_{0}\alpha}\nonumber \\
 & \qquad+s_{2}\Vert\boldsymbol{\xi}(\boldsymbol{x})\Vert_{\bullet}^{bs_{0}}-\tilde{\lambda}s_{2}\lVert\boldsymbol{\xi}(\boldsymbol{x})\rVert_{\bullet}^{bs_{0}}+C.\label{Step5_1}
\end{align}
To combine the terms involving $|z_{1}|^{2s_{0}\alpha}$, set $\bar{r}=\tilde{r}-\tilde{s}_{1}$
so that 
\[
-\tilde{r}|z_{1}|^{2s_{0}\alpha}+\tilde{s}_{1}|z_{1}|^{2s_{0}\alpha}=-\bar{r}|z_{1}|^{2s_{0}\alpha};
\]
recalling that\textcolor{black}{{} $\tilde{s}_{1}$ can be chosen as
close to zero as desired}, we have $\bar{r}>0$ by a suitable choice
of\textcolor{black}{{} $\tilde{s}_{1}$.} On the other hand, to manipulate
the terms involving $\lVert\boldsymbol{\xi}(\boldsymbol{x})\rVert_{\bullet}^{bs_{0}}$,
choose $s_{2}$ large enough to ensure that $\tilde{\lambda}-C/s_{2}\in(0,1)$
and set $\bar{\lambda}=\tilde{\lambda}-C/s_{2}$. As now $-\bar{\lambda}s_{2}=C-\tilde{\lambda}s_{2}$,
the terms involving $\lVert\boldsymbol{\xi}(\boldsymbol{x})\rVert_{\bullet}^{bs_{0}}$
in (\ref{Step5_1}) can be written as
\begin{equation}
s_{2}\Vert\boldsymbol{\xi}(\boldsymbol{x})\Vert_{\bullet}^{bs_{0}}+C\Vert\boldsymbol{\xi}(\boldsymbol{x})\Vert_{\bullet}^{bs_{0}}-\tilde{\lambda}s_{2}\Vert\boldsymbol{\xi}(\boldsymbol{x})\Vert_{\bullet}^{bs_{0}}=s_{2}\Vert\boldsymbol{\xi}(\boldsymbol{x})\Vert_{\bullet}^{bs_{0}}-\bar{\lambda}s_{2}\Vert\boldsymbol{\xi}(\boldsymbol{x})\Vert_{\bullet}^{bs_{0}}.\label{Step5_2}
\end{equation}
Whenever $s_{2}\Vert\boldsymbol{\xi}(\boldsymbol{x})\Vert_{\bullet}^{bs_{0}}<1$,
(\ref{Step5_2}) is bounded by the constant $1$; for $s_{2}\Vert\boldsymbol{\xi}(\boldsymbol{x})\Vert_{\bullet}^{bs_{0}}\geq1$,
we have $s_{2}\Vert\boldsymbol{\xi}(\boldsymbol{x})\Vert_{\bullet}^{bs_{0}}\geq s_{2}^{\alpha}\Vert\boldsymbol{\xi}(\boldsymbol{x})\Vert_{\bullet}^{bs_{0}\alpha}$
as $\alpha\in(0,1)$. Thus the expression in (\ref{Step5_2}) is always
bounded by $1+s_{2}\Vert\boldsymbol{\xi}(\boldsymbol{x})\Vert_{\bullet}^{bs_{0}}-\bar{\lambda}s_{2}^{\alpha}\Vert\boldsymbol{\xi}(\boldsymbol{x})\Vert_{\bullet}^{bs_{0}\alpha}$.
As $V(\boldsymbol{x})=1+|z_{1}|^{2s_{0}}+s_{1}\left\Vert \boldsymbol{z}_{2}\right\Vert _{*}^{2s_{0}\alpha}+s_{2}\Vert\boldsymbol{\xi}(\boldsymbol{x})\Vert_{\bullet}^{bs_{0}}$,
we obtain from (\ref{Step5_1}) that 
\begin{equation}
E[V(\boldsymbol{y}_{1})\mid\boldsymbol{y}_{0}=\boldsymbol{x}]\leq V(\boldsymbol{x})-(1+\bar{r}|z_{1}|^{2s_{0}\alpha}+\tilde{\varpi}s_{1}^{\alpha}\left\Vert \boldsymbol{z}_{2}\right\Vert _{*}^{2s_{0}\alpha^{2}}+\bar{\lambda}s_{2}^{\alpha}\Vert\boldsymbol{\xi}(\boldsymbol{x})\Vert_{\bullet}^{bs_{0}\alpha})+C.\label{Step5_new}
\end{equation}
The inequality in (\ref{c_r-ineq}) implies that
\[
[V(\boldsymbol{x})]^{\alpha}=(1+|z_{1}|^{2s_{0}}+s_{1}\left\Vert \boldsymbol{z}_{2}\right\Vert _{*}^{2s_{0}\alpha}+s_{2}\Vert\boldsymbol{\xi}(\boldsymbol{x})\Vert_{\bullet}^{bs_{0}})^{\alpha}\leq1+|z_{1}|^{2s_{0}\alpha}+s_{1}^{\alpha}\left\Vert \boldsymbol{z}_{2}\right\Vert _{*}^{2s_{0}\alpha^{2}}+s_{2}^{\alpha}\Vert\boldsymbol{\xi}(\boldsymbol{x})\Vert_{\bullet}^{bs_{0}\alpha}
\]
so that, setting $e=\min\{\bar{r},\tilde{\varpi},\bar{\lambda}\}\in(0,1)$,
we have
\[
e[V(\boldsymbol{x})]^{\alpha}\leq1+\bar{r}|z_{1}|^{2s_{0}\alpha}+\tilde{\varpi}s_{1}^{\alpha}\left\Vert \boldsymbol{z}_{2}\right\Vert _{*}^{2s_{0}\alpha^{2}}+\bar{\lambda}s_{2}^{\alpha}\Vert\boldsymbol{\xi}(\boldsymbol{x})\Vert_{\bullet}^{bs_{0}\alpha}.
\]
Therefore, setting $\tilde{e}=e/2$,
\[
E[V(\boldsymbol{y}_{1})\mid\boldsymbol{y}_{0}=\boldsymbol{x}]\leq V(\boldsymbol{x})-\tilde{e}[V(\boldsymbol{x})]^{\alpha}+\left\{ C-\tilde{e}[V(\boldsymbol{x})]^{\alpha}\right\} .
\]

Now, define the set
\begin{equation}
A_{N}=\bigl\{\boldsymbol{x}\in\mathbb{R}^{p+q}:|z_{1}(\boldsymbol{x}_{1})|^{2s_{0}}\leq N,\quad\Vert\boldsymbol{z}_{2}(\boldsymbol{x}_{1})\Vert_{*}^{2s_{0}\alpha}\leq N,\quad\Vert\boldsymbol{\xi}(\boldsymbol{x})\Vert_{\bullet}^{bs_{0}\alpha}\leq N\bigr\},\label{Step5_Set_AN}
\end{equation}
where $N$ is so large that $A_{N}$ is nonempty (see (\ref{ARCH(q)})).
The complement of $A_{N}$ is denoted by $A_{N}^{c}$ so that $\boldsymbol{x}\in A_{N}^{c}$
if either $|z_{1}(\boldsymbol{x}_{1})|^{2s_{0}}>N$, $\Vert\boldsymbol{z}_{2}(\boldsymbol{x}_{1})\Vert_{*}^{2s_{0}\alpha}>N$
or $\Vert\boldsymbol{\xi}(\boldsymbol{x})\Vert_{\bullet}^{bs_{0}\alpha}>N$.
By choosing a large enough $N$, for all $\boldsymbol{x}\in A_{N}^{c}$
it holds that $C-\tilde{e}[V(\boldsymbol{x})]^{\alpha}<0$ so that
$E[V(\boldsymbol{y}_{1})\mid\boldsymbol{y}_{0}=\boldsymbol{x}]\leq V(\boldsymbol{x})-\tilde{e}[V(\boldsymbol{x})]^{\alpha}$
for all $\boldsymbol{x}\in A_{N}^{c}$. On the other hand, the function
$V(\boldsymbol{x})-e[V(\boldsymbol{x})]^{\alpha}+C$ is clearly bounded
by some positive constant $\tilde{b}$ on $A_{N}$. Therefore we can
conclude that there exists an $N$ and a positive constant $\tilde{b}$
such that

\begin{equation}
E[V(\boldsymbol{y}_{1})\mid\boldsymbol{y}_{0}=\boldsymbol{x}]\leq V(\boldsymbol{x})-\phi_{1}\left(V(\boldsymbol{x})\right)+\tilde{b}\boldsymbol{1}_{A_{N}}(\boldsymbol{x}),\label{Step5_End}
\end{equation}
where $\phi_{1}(v)=\tilde{e}v{}^{\alpha}$. This implies that Condition
D holds with $\phi=\phi_{1}$ and $C=A_{N}$.

\paragraph*{Step 6: Showing that $A_{N}$ is petite.}

\noindent We first note that the definition of a petite set and other
Markov chain concepts we refer to below can be found in \citet[Chs 4--6]{meyn2009markov}.
The idea is to establish that the (potentially non-compact) set $A_{N}$
in (\ref{Step5_Set_AN}) is petite for any $N\geq1$ so large that
$A_{N}$ is nonempty. To this end, we show below that there exists
an $M_{N}<\infty$ such that
\begin{equation}
\sup_{\boldsymbol{x}\in A_{N}}E[\Vert\boldsymbol{y}_{p+q}\Vert_{1}^{2s_{0}\alpha}\mid\boldsymbol{y}_{0}=\boldsymbol{x}]<M_{N}^{2s_{0}\alpha},\label{Step6_1}
\end{equation}
where $\Vert\cdot\Vert_{1}$ denotes the usual $l_{1}$ vector norm.
Next note that Theorem 2.2(ii) of \citet{cline1998verifying} along
with our Assumption 1 shows that the Markov chain $\boldsymbol{y}_{t}$
is a $\psi$-irreducible and aperiodic T-chain (see also Example 2.1
of the aforementioned paper). Therefore the compact set $B_{N}=\{\boldsymbol{x}\in\mathbb{R}^{p+q}:\Vert\boldsymbol{x}\Vert_{1}\leq M_{N}\}$
is small (see \citealp[Thms 6.2.5(ii) and 5.5.7]{meyn2009markov}).
Moreover, due to Markov's inequality, 
\begin{align*}
\inf_{\boldsymbol{x}\in A_{N}}\Pr[\boldsymbol{y}_{p+q}\in B_{N}\mid\boldsymbol{y}_{0}=\boldsymbol{x}] & =1-\sup_{\boldsymbol{x}\in A_{N}}\Pr[\Vert\boldsymbol{y}_{p+q}\Vert_{1}\geq M_{N}\mid\boldsymbol{y}_{0}=\boldsymbol{x}]\\
 & \geq1-\sup_{\boldsymbol{x}\in A_{N}}E[\Vert\boldsymbol{y}_{p+q}\Vert_{1}^{2s_{0}\alpha}\mid\boldsymbol{y}_{0}=\boldsymbol{x}]/M_{N}^{2s_{0}\alpha},
\end{align*}
where the last expression is positive due to (\ref{Step6_1}). Proposition
5.2.4(i) of \citet{meyn2009markov} now implies that the set $A_{N}$
is small. Proposition 5.5.3 of the same reference therefore implies
that the set $A_{N}$ is also petite.

To complete the proof of petiteness of $A_{N}$, it remains to establish
(\ref{Step6_1}). First we introduce some notation. We let $\mnorm{\,\cdot\,}_{1}$
denote the maximum column sum norm defined for real square matrices
(this norm is induced by the $l_{1}$ vector norm, see \citealp[Sec 5.6]{HornJohnson2013}).
For brevity, we denote $\boldsymbol{z}_{t}=\boldsymbol{z}(\boldsymbol{y}_{1,t})=\mathbf{A}\boldsymbol{y}_{1,t}$
and also partition the $p$-dimensional $\boldsymbol{z}_{t}$ as $\boldsymbol{z}_{t}=(z_{1,t},\boldsymbol{z}_{2,t})$
(see (\ref{z})). This allows us to write the companion form (\ref{Companion form_A2})
as
\begin{equation}
\boldsymbol{z}_{t}=\begin{bmatrix}z_{1,t}\\
\boldsymbol{z}_{2,t}
\end{bmatrix}=\begin{bmatrix}g(z_{1,t-1})+\sigma_{t}\varepsilon_{t}\\
\boldsymbol{\Pi}_{1}\boldsymbol{z}_{2,t-1}+z_{1,t-1}\boldsymbol{\iota}_{p-1}
\end{bmatrix}.\label{Companion form_A4}
\end{equation}
Finally, we set $\vec{\boldsymbol{z}}_{1,p+q}=(z_{1,p+q},\ldots,z_{1,1})$
and $\vec{\boldsymbol{z}}_{2,p+q}=(\boldsymbol{z}_{2,p+q},\ldots,\boldsymbol{z}_{2,1})$.

Now consider the norm $\Vert\boldsymbol{y}_{p+q}\Vert_{1}$ in (\ref{Step6_1}).
Using properties of the norms $\Vert\cdot\Vert_{1}$ and $\mnorm{\,\cdot\,}_{1}$
we can write 
\[
\lVert\boldsymbol{y}_{p+q}\rVert_{1}\leq\sum_{t=1}^{p+q}\lVert\boldsymbol{y}_{1,t}\rVert_{1}=\sum_{t=1}^{p+q}\lVert\boldsymbol{A}^{-1}\boldsymbol{z}_{t}\rVert_{1}\leq\mnorm{\boldsymbol{A}^{-1}}_{1}\sum_{t=1}^{p+q}\lVert\boldsymbol{z}_{t}\rVert_{1}=\mnorm{\boldsymbol{A}^{-1}}_{1}(\lVert\vec{\boldsymbol{z}}_{1,p+q}\rVert_{1}+\lVert\vec{\boldsymbol{z}}_{2,p+q}\rVert_{1})
\]
implying that $\lVert\boldsymbol{y}_{p+q}\rVert_{1}\leq C(\lVert\vec{\boldsymbol{z}}_{1,p+q}\rVert_{1}+\lVert\vec{\boldsymbol{z}}_{2,p+q}\rVert_{1})$.
Adding terms and making use of inequality (\ref{c_r-ineq}) and the
fact that $\alpha\in(0,1)$ we obtain
\begin{align}
\lVert\boldsymbol{y}_{p+q}\rVert_{1}^{2s_{0}\alpha} & \leq C((1+\lVert\vec{\boldsymbol{z}}_{1,p+q}\rVert_{1})^{2s_{0}\alpha}+\lVert\vec{\boldsymbol{z}}_{2,p+q}\rVert_{1}^{2s_{0}\alpha})\nonumber \\
 & \leq C((1+\lVert\vec{\boldsymbol{z}}_{1,p+q}\rVert_{1})^{2s_{0}}+\lVert\vec{\boldsymbol{z}}_{2,p+q}\rVert_{1}^{2s_{0}\alpha})\nonumber \\
 & \leq C(1+\lVert\vec{\boldsymbol{z}}_{1,p+q}\rVert_{1}^{2s_{0}}+\lVert\vec{\boldsymbol{z}}_{2,p+q}\rVert_{1}^{2s_{0}\alpha}).\label{Step6_Bound1}
\end{align}

To obtain an upper bound for the term $\lVert\vec{\boldsymbol{z}}_{1,p+q}\rVert_{1}^{2s_{0}}$,
consider the equality $z_{1,t}=g(z_{1,t-1})+\sigma_{t}\varepsilon_{t}$
from (\ref{Companion form_A4}) and note that, by Assumption 2, $|g(u)|\leq K_{0}$
for $\left|u\right|\leq M_{0}$ and $|g(u)|\leq(1-r\left|u\right|^{-\rho})|u|\leq|u|$
for $\left|u\right|\geq M_{0}$, so that $|g(u)|\leq K_{0}+|u|$ for
all $u\in\mathbb{R}$ (note that Assumption 2 requires $M_{0}$ to
be so large that $r\left|u\right|^{-\rho}\in(0,1)$ for $\left|u\right|\geq M_{0}$).
Using these inequalities, $|z_{1,t}|\leq K_{0}+|z_{1,t-1}|+\sigma_{t}|\varepsilon_{t}|$
($t=1,\ldots,p+q$), so that
\begin{align*}
|z_{1,1}| & \leq K_{0}+|z_{1,0}|+\sigma_{1}|\varepsilon_{1}|,\\
|z_{1,2}| & \leq2K_{0}+|z_{1,0}|+\sigma_{1}|\varepsilon_{1}|+\sigma_{2}|\varepsilon_{2}|,\\
 & \vdots\\
|z_{1,p+q}| & \leq(p+q)K_{0}+|z_{1,0}|+\sigma_{1}|\varepsilon_{1}|+\cdots+\sigma_{p+q}|\varepsilon_{p+q}|.
\end{align*}
Thus $\Vert\vec{\boldsymbol{z}}_{1,p+q}\Vert_{1}\leq C\left(1+|z_{1,0}|+\sum_{i=1}^{p+q}\sigma_{i}|\varepsilon_{i}|\right)$
and, making use of inequality (\ref{c_r-ineq}), 
\begin{equation}
\Vert\vec{\boldsymbol{z}}_{1,p+q}\Vert_{1}^{2s_{0}}\leq C\biggl(1+|z_{1,0}|^{2s_{0}}+\sum_{i=1}^{p+q}\sigma_{i}^{2s_{0}}|\varepsilon_{i}|^{2s_{0}}\biggr).\label{Step6_Bound2}
\end{equation}

Next, to bound the term $\lVert\vec{\boldsymbol{z}}_{2,p+q}\rVert_{1}^{2s_{0}\alpha}$,
consider $\boldsymbol{z}_{2,t}=\boldsymbol{\Pi}_{1}\boldsymbol{z}_{2,t-1}+z_{1,t-1}\boldsymbol{\iota}_{p-1}$
(see (\ref{Companion form_A4})). Setting $\kappa=\mnorm{\boldsymbol{\Pi}_{1}}_{1}$
and using the fact $\lVert\boldsymbol{\iota}_{p-1}\rVert_{1}=1$ we
obtain
\[
\lVert\boldsymbol{z}_{2,t}\rVert_{1}\leq\mnorm{\boldsymbol{\Pi}_{1}}_{1}\lVert\boldsymbol{z}_{2,t-1}\rVert_{1}+|z_{1,t-1}|\lVert\boldsymbol{\iota}_{p-1}\rVert_{1}=\kappa\lVert\boldsymbol{z}_{2,t-1}\rVert_{1}+|z_{1,t-1}|
\]
and furthermore
\begin{align*}
\lVert\boldsymbol{z}_{2,1}\rVert_{1} & \leq\kappa\lVert\boldsymbol{z}_{2,0}\rVert_{1}+|z_{1,0}|,\\
\lVert\boldsymbol{z}_{2,2}\rVert_{1} & \leq\kappa^{2}\lVert\boldsymbol{z}_{2,0}\rVert_{1}+\kappa|z_{1,0}|+|z_{1,1}|,\\
 & \vdots\\
\lVert\boldsymbol{z}_{2,p+q}\rVert_{1} & \leq\kappa^{p+q}\lVert\boldsymbol{z}_{2,0}\rVert_{1}+\kappa^{p+q-1}|z_{1,0}|+\cdots+|z_{1,p+q-1}|.
\end{align*}
This implies that 
\[
\lVert\vec{\boldsymbol{z}}_{2,p+q}\rVert_{1}\leq C(\lVert\boldsymbol{z}_{2,0}\rVert_{1}+1+|z_{1,0}|+\lVert\vec{\boldsymbol{z}}_{1,p+q}\rVert_{1}).
\]
As the norms $\lVert\cdot\rVert_{1}$ and $\lVert\cdot\rVert_{*}$
are equivalent, it holds that $\lVert\boldsymbol{z}_{2,0}\rVert_{1}\leq C\lVert\boldsymbol{z}_{2,0}\rVert_{*}$.
Making use of inequality (\ref{c_r-ineq}) and the fact that $\alpha\in(0,1)$
we obtain
\begin{align}
\lVert\vec{\boldsymbol{z}}_{2,p+q}\rVert_{1}^{2s_{0}\alpha} & \leq C(\lVert\boldsymbol{z}_{2,0}\rVert_{*}^{2s_{0}\alpha}+(1+|z_{1,0}|+\lVert\vec{\boldsymbol{z}}_{1,p+q}\rVert_{1})^{2s_{0}\alpha})\nonumber \\
 & \leq C(\lVert\boldsymbol{z}_{2,0}\rVert_{*}^{2s_{0}\alpha}+(1+|z_{1,0}|+\lVert\vec{\boldsymbol{z}}_{1,p+q}\rVert_{1})^{2s_{0}})\nonumber \\
 & \leq C(1+\lVert\boldsymbol{z}_{2,0}\rVert_{*}^{2s_{0}\alpha}+|z_{1,0}|^{2s_{0}}+\lVert\vec{\boldsymbol{z}}_{1,p+q}\rVert_{1}^{2s_{0}}).\label{Step6_Bound3}
\end{align}

Now combine (\ref{Step6_Bound1}) with the upper bounds obtained for
$\lVert\vec{\boldsymbol{z}}_{1,p+q}\rVert_{1}^{2s_{0}}$ and $\lVert\vec{\boldsymbol{z}}_{2,p+q}\rVert_{1}^{2s_{0}\alpha}$
in (\ref{Step6_Bound2}) and (\ref{Step6_Bound3}), and recall that
$z_{1,0}=z_{1}(\boldsymbol{y}_{1,0})$ and $\boldsymbol{z}_{2,0}=\boldsymbol{z}_{2}(\boldsymbol{y}_{1,0})$,
to obtain
\[
\lVert\boldsymbol{y}_{p+q}\rVert_{1}^{2s_{0}\alpha}\leq C\biggl(1+\lVert\boldsymbol{z}_{2}(\boldsymbol{y}_{1,0})\rVert_{*}^{2s_{0}\alpha}+|z_{1}(\boldsymbol{y}_{1,0})|^{2s_{0}}+\sum_{i=1}^{p+q}\sigma_{i}^{2s_{0}}|\varepsilon_{i}|^{2s_{0}}\biggr).
\]
As $\mu_{2s_{0}}=E[|\varepsilon_{1}|^{2s_{0}}]$ is finite, this implies
that
\begin{equation}
E[\Vert\boldsymbol{y}_{p+q}\Vert_{1}^{2s_{0}\alpha}\mid\boldsymbol{y}_{0}=\boldsymbol{x}]\leq C\biggl(1+\lVert\boldsymbol{z}_{2}(\boldsymbol{x}_{1})\rVert_{*}^{2s_{0}\alpha}+|z_{1}(\boldsymbol{x}_{1})|^{2s_{0}}+\mu_{2s_{0}}\sum_{i=1}^{p+q}E[\sigma_{i}^{2s_{0}}\mid\boldsymbol{y}_{0}=\boldsymbol{x}]\biggr).\hspace{-5pt}\label{Step6_Bound4}
\end{equation}

Next consider the terms in (\ref{Step6_Bound4}) involving conditional
expectations of the $\sigma_{i}^{2s_{0}}$'s. We first derive an inequality
which is similar to inequality (11) in \citet{meitz2010SPL}. Using
repeated substitution and the equality $\boldsymbol{\xi}(\boldsymbol{y}_{t})=\Lambda_{\zeta,t}\boldsymbol{\xi}(\boldsymbol{y}_{t-1})+\boldsymbol{\omega}_{\zeta,t}$
we obtain, for any fixed $t\geq1$, that
\[
\boldsymbol{\xi}(\boldsymbol{y}_{t})=\prod_{k=0}^{t-1}\Lambda_{\zeta,t-k}\boldsymbol{\xi}(\boldsymbol{y}_{0})+\boldsymbol{\omega}_{\zeta,t}+\sum_{k=0}^{t-2}\prod_{l=0}^{k}\Lambda_{\zeta,t-l}\boldsymbol{\omega}_{\zeta,t-k-1}.
\]
Now consider the vector norm $\Vert\cdot\Vert_{\bullet}$ in Assumption
4. Denote by $\mnorm{\,\cdot\,}_{\bullet}$ the matrix norm induced
by the vector norm $\Vert\cdot\Vert_{\bullet}$; that is, for any
$q\times q$ matrix $A$, set
\[
\mnorm{A}_{\bullet}=\max_{\Vert\boldsymbol{x}\Vert_{\bullet}=1}\Vert A\boldsymbol{x}\Vert_{\bullet}\qquad(\boldsymbol{x}\in\mathbb{R}^{q}).
\]
(For clarity, note that $\mnorm{\,\cdot\,}_{\bullet}$ above and $\mnorm{\,\cdot\,}_{\bullet L^{p}}$
defined in (\ref{Matrix norm_definition}) coincide for nonrandom
matrices but differ for random ones.) As $\Vert\cdot\Vert_{\bullet}$
in Assumption 4 is assumed to be monotone, it follows from Problems
5.6.P41(c) and 5.6.P42 in \citet[p. 368]{HornJohnson2013} that the
induced matrix norm $\mnorm{\,\cdot\,}_{\bullet}$ is monotone on
the positive orthant, meaning that any $q\times q$ matrices $A$
and $B$ that satisfy the (entrywise) inequalities $0\leq A\leq B$
also satisfy the inequality $\mnorm{A}_{\bullet}\leq\mnorm{B}_{\bullet}$.
Usual properties of vector norms and matrix norms in conjunction with
inequality (\ref{c_r-ineq}) therefore yield
\[
\Vert\boldsymbol{\xi}(\boldsymbol{y}_{t})\Vert_{\bullet}^{s_{0}}\leq C\prod_{k=0}^{t-1}\mnorm{\Lambda_{\zeta,t-k}}_{\bullet}^{s_{0}}\Vert\boldsymbol{\xi}(\boldsymbol{y}_{0})\Vert_{\bullet}^{s_{0}}+C\Vert\boldsymbol{\omega}_{\zeta,t}\Vert_{\bullet}^{s_{0}}+C\sum_{k=0}^{t-2}\prod_{l=0}^{k}\mnorm{\Lambda_{\zeta,t-l}}_{\bullet}^{s_{0}}\Vert\boldsymbol{\omega}_{\zeta,t-k-1}\Vert_{\bullet}^{s_{0}}.
\]
By the monotonicity properties of the norms $\Vert\cdot\Vert_{\bullet}$
and $\mnorm{\,\cdot\,}_{\bullet}$ and the definitions of the matrices
$\Lambda_{\zeta,t}$ and $\Lambda_{t}$ in (\ref{Companion form_A3})
and (\ref{Equations Lambda}) we also obtain $\mnorm{\Lambda_{\zeta,t}}_{\bullet}\leq\mnorm{\Lambda_{t}}_{\bullet}$
and $\lVert\boldsymbol{\omega}_{\zeta,t}\rVert_{\bullet}\leq\lVert\boldsymbol{\omega}_{t}\rVert_{\bullet}=\varepsilon_{t}^{2}\lVert\boldsymbol{\omega}\rVert_{\bullet}$
(a.s.) for all $t=1,2,\ldots$, implying that 
\[
\Vert\boldsymbol{\xi}(\boldsymbol{y}_{t})\Vert_{\bullet}^{s_{0}}\leq C\prod_{k=0}^{t-1}\mnorm{\Lambda_{t-k}}_{\bullet}^{s_{0}}\Vert\boldsymbol{\xi}(\boldsymbol{y}_{0})\Vert_{\bullet}^{s_{0}}+C\biggl(|\varepsilon_{t}|^{2s_{0}}+\sum_{k=0}^{t-2}\prod_{l=0}^{k}\mnorm{\Lambda_{t-l}}_{\bullet}^{s_{0}}|\varepsilon_{t-k-1}|^{2s_{0}}\biggr)\Vert\boldsymbol{\omega}\Vert_{\bullet}^{s_{0}}.
\]

Now, denote the expectation $E\left[\mnorm{\Lambda_{t}}_{\bullet}^{s_{0}}\right]$
by $\chi$ (this expectation is finite due to Assumption 4). Using
the independence of the $\Lambda_{t}$'s and independence of $\mnorm{\Lambda_{t-l}}_{\bullet}$'s
and $\varepsilon_{t-k-1}$'s, yields
\begin{equation}
E\left[\Vert\boldsymbol{\xi}(\boldsymbol{y}_{t})\Vert_{\bullet}^{s_{0}}\mid\boldsymbol{y}_{0}=\boldsymbol{x}\right]\leq C\chi^{t}\Vert\boldsymbol{\xi}(\boldsymbol{x})\Vert_{\bullet}^{s_{0}}+C\biggl(1+\sum_{k=0}^{t-2}\chi^{k+1}\biggr)\Vert\boldsymbol{\omega}\Vert_{\bullet}^{s_{0}}.\label{Extended ineq ksi(y)}
\end{equation}

\noindent Inequality (\ref{Ineq cond var}) in conjunction with (\ref{c_r-ineq})
show that $\sigma_{i}^{2s_{0}}\leq C(1+\Vert\boldsymbol{\xi}(\boldsymbol{y}_{i-1})\Vert_{\bullet}^{s_{0}})$
(a.s.) for all $i=1,\ldots,p+q$. From (\ref{Extended ineq ksi(y)})
it then follows that $E[\sigma_{i}^{2s_{0}}\mid\boldsymbol{y}_{0}=\boldsymbol{x}]\leq C(1+\Vert\boldsymbol{\xi}(\boldsymbol{x})\Vert_{\bullet}^{s_{0}})$
which together with (\ref{Step6_Bound4}) implies 
\[
E[\Vert\boldsymbol{y}_{p+q}\Vert_{1}^{2s_{0}\alpha}\mid\boldsymbol{y}_{0}=\boldsymbol{x}]\leq C\left(1+|z_{1}(\boldsymbol{x}_{1})|^{2s_{0}}+\lVert z_{2}(\boldsymbol{x}_{1})\rVert_{*}^{2s_{0}\alpha}+\Vert\boldsymbol{\xi}(\boldsymbol{x})\Vert_{\bullet}^{s_{0}}\right).
\]
For any $\boldsymbol{x}\in A_{N}$, the dominant side is bounded by
$C(1+2N+N^{1/b\alpha})$ and thus we can find a finite $M_{N}$ such
that (\ref{Step6_1}) holds.

\paragraph*{Step 7: Completing the proof.}

\noindent We are now ready to complete the proof by applying Theorem
1(iii) in \citet{meitz2022subgear}. To this end, in the beginning
of Step 6 we already noted that the Markov chain $\boldsymbol{y}_{t}$
is $\psi$-irreducible and aperiodic. That Condition D holds was shown
in (\ref{Step5_End}) in Step 5. Petiteness of the set $A_{N}$ was
shown in Step 6. We also need to verify that $\sup_{\boldsymbol{x}\in A_{N}}V(\boldsymbol{x})<\infty$;
this inequality is a straightforward consequence of the definitions
of the set $A_{N}$ and the function $V$. Thus, applying Theorem
1(iii) in \citet{meitz2022subgear} we can complete the proof. \qed

\bigskip{}

\begin{proof}[\textbf{\emph{Details for the finiteness of moments in Section 3.2}}]
 The arguments are similar to those used in the proof of Corollary
to Theorem 3 in \citet{meitz2022subgear}. First note that inequality
(\ref{Step3_Final}) continues to hold if the term $\tilde{\varpi}s_{1}^{\alpha}\left\Vert \boldsymbol{z}_{2}\right\Vert _{*}^{2s_{0}\alpha^{2}}$
on its dominant side is replaced with the term $\tilde{\varpi}s_{1}\left\Vert \boldsymbol{z}_{2}\right\Vert _{*}^{2s_{0}\alpha}$
(this can be seen from (\ref{Step3_2}) and the arguments that follow
it). Consequently, the same replacement can be done on the dominant
sides of inequalities (\ref{Step5_1}) and (\ref{Step5_new}), the
latter inequality thus becoming
\[
E[V(\boldsymbol{y}_{1})\mid\boldsymbol{y}_{0}=\boldsymbol{x}]\leq V(\boldsymbol{x})-(1+\bar{r}|z_{1}(\boldsymbol{x}_{1})|^{2s_{0}\alpha}+\tilde{\varpi}s_{1}\left\Vert \boldsymbol{z}_{2}(\boldsymbol{x}_{1})\right\Vert _{*}^{2s_{0}\alpha}+\bar{\lambda}s_{2}^{\alpha}\Vert\boldsymbol{\xi}(\boldsymbol{x})\Vert_{\bullet}^{bs_{0}\alpha})+C.
\]
Finiteness of certain moments with respect to the stationary distribution
$\pi$ of $\boldsymbol{y}_{t}$ can now be obtained from Theorem 14.3.7
of \citet{meyn2009markov}, namely  $\int_{\mathbb{R}^{p+q}}(1+\bar{r}|z_{1}(\boldsymbol{x}_{1})|^{2s_{0}\alpha}+\tilde{\varpi}s_{1}\left\Vert \boldsymbol{z}_{2}(\boldsymbol{x}_{1})\right\Vert _{*}^{2s_{0}\alpha}+\bar{\lambda}s_{2}^{\alpha}\Vert\boldsymbol{\xi}(\boldsymbol{x})\Vert_{\bullet}^{bs_{0}\alpha})\pi(d\boldsymbol{x})<\infty$.
Noting that $2s_{0}\alpha=2s_{0}-\rho$ and following the arguments
in the proof of Corollary to Theorem 3 in \citet{meitz2022subgear}
it follows that the stationary version of $\boldsymbol{y}_{t}$ satisfies
$E[|y_{t}|^{2s_{0}-\rho}]<\infty$. 
\end{proof}
\bigskip{}

\begin{proof}[\textbf{\emph{Proofs of Propositions 1 and 2}}]
 For Proposition 1, note that model (\ref{LSTAR model}) can be written
as $u_{t}=u_{t-1}+\nu_{1}L(u_{t-1};\gamma,a_{1})+\nu_{2}(1-L(u_{t-1};\gamma,a_{2}))+\sigma_{t}\varepsilon_{t}$
so that the function $g(\cdot)$ in Assumption 2(ii) takes the form
$g(u)=u+\nu_{1}L(u;\gamma,a_{1})+\nu_{2}(1-L(u;\gamma,a_{2}))$. Arguments
used in the proof of Proposition 1 in \citet{meitz2022subgear} now
show that Assumption 2(ii) holds with $\rho=1$. Applying Theorem
1 with $\delta=2s_{0}$ yields the polynomial ergodicity result, and
the moment result follows from the remarks made after Theorem 1. As
for Proposition 2, model (\ref{ESTAR model}) can be written as $u_{t}=S(u_{t-1})u_{t-1}+\sigma_{t}\varepsilon_{t}$
so that now $g(u)=S(u)u$. Assumption 2(ii) can be verified as in
the proof of Proposition 2 in \citet{meitz2022subgear}, and the result
follows from Theorem 1 (with $\delta=2s_{0}/\rho$).
\end{proof}
\pagebreak{}

\section*{Appendix B}

Appendix B contains Figure 2 which displays further analysis of the
residuals of model (\ref{EstimationResults}).

\begin{figure}[H]
\vspace*{-20pt}

\hfill{}%
\begin{minipage}[t][1\totalheight][c]{0.45\textwidth}%
\includegraphics[width=1.1\textwidth]{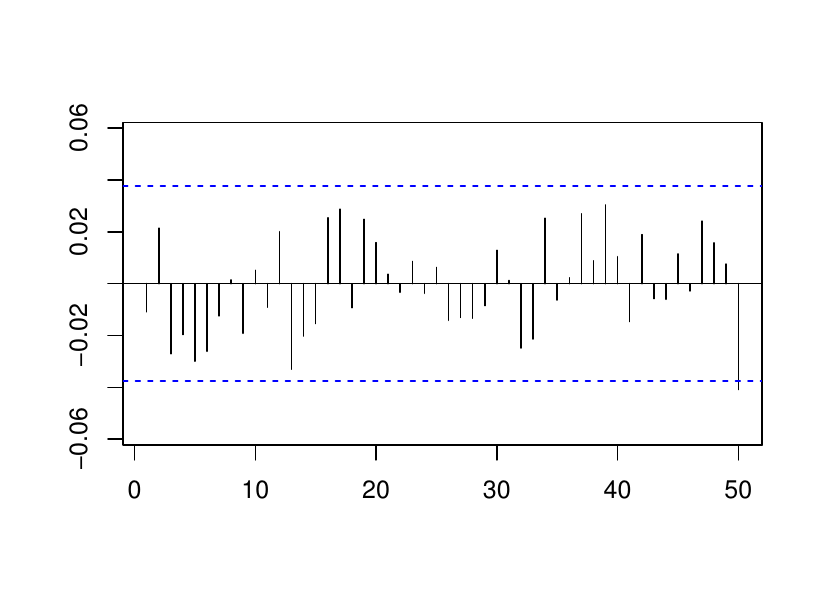}%
\end{minipage}\hfill{}%
\begin{minipage}[t][1\totalheight][c]{0.45\textwidth}%
\includegraphics[width=1.1\textwidth]{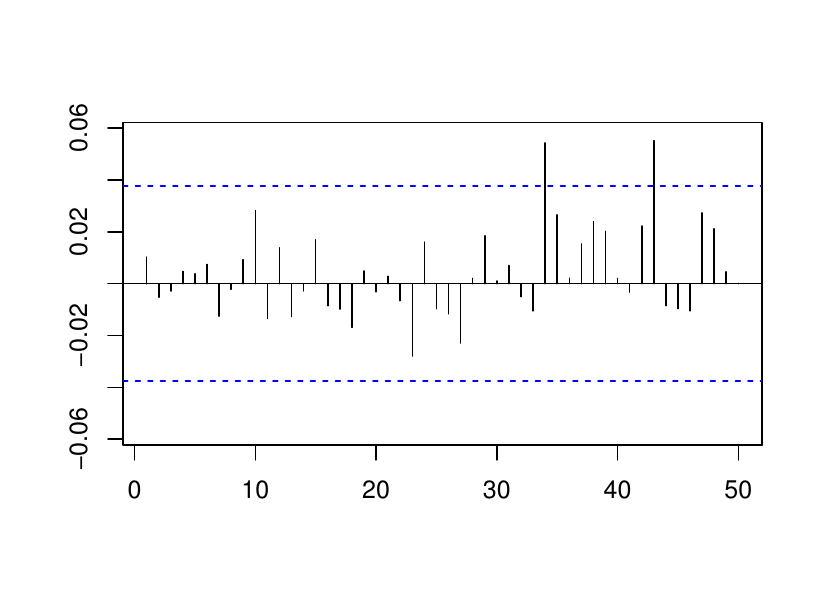}%
\end{minipage}\hfill{}\vspace*{-50pt}

\hfill{}%
\begin{minipage}[t][1\totalheight][c]{0.45\textwidth}%
\includegraphics[width=1.1\textwidth]{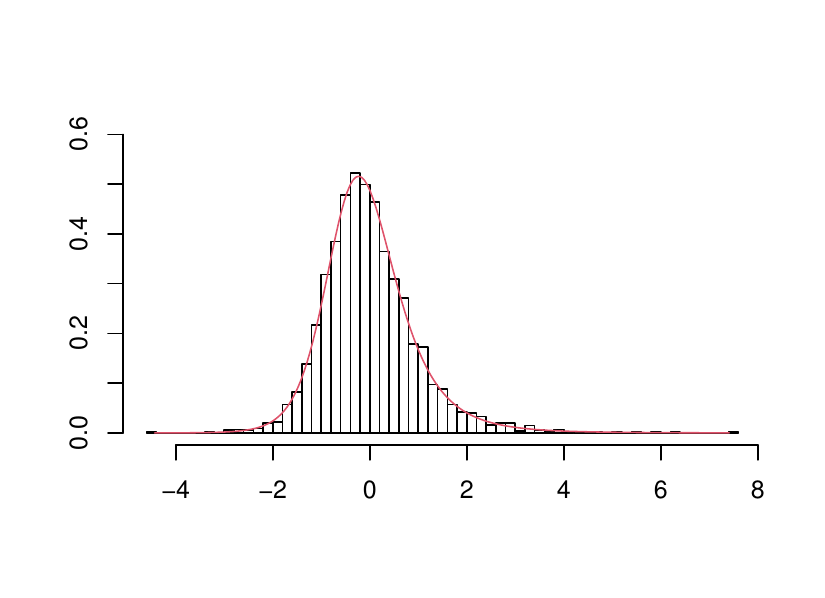}%
\end{minipage}\hfill{}%
\begin{minipage}[t][1\totalheight][c]{0.45\textwidth}%
\includegraphics[width=1.1\textwidth]{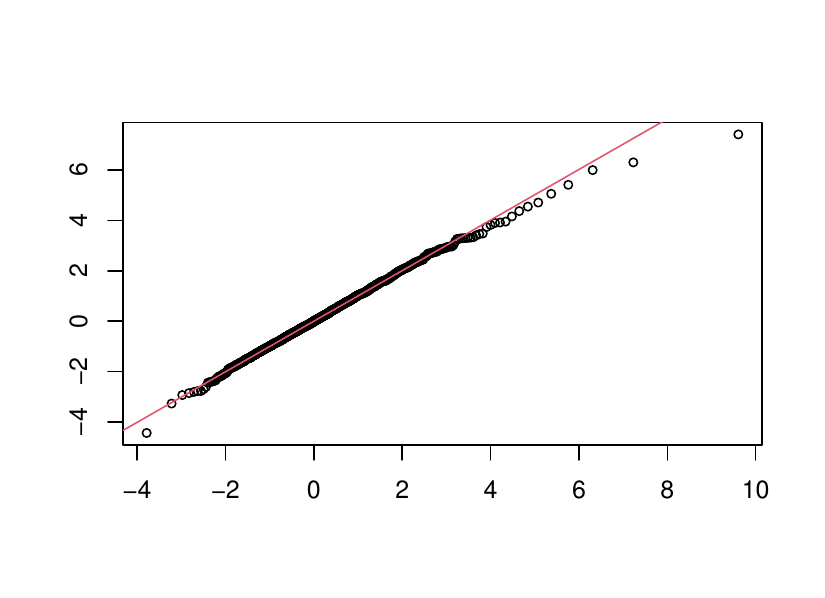}%
\end{minipage}\hfill{}\vspace*{-25pt}
\caption{\label{FigEx2} Further analysis of the residuals shown in (the bottom
right graph of) Figure 1: autocorrelation function of $\hat{\varepsilon}_{t}$
(top left), autocorrelation function of $\hat{\varepsilon}_{t}^{2}$
(top right), histogram along with the estimated error density (bottom
left), and a Q-Q plot (bottom right). The dashed lines in the autocorrelation
function graphs show the conventional bounds $\pm1.96/\sqrt{T}\approx\pm0.038$
($T=2715$; the first four observations are used as initial values). }
\end{figure}
\smallskip{}

\bibliographystyle{chicago}
\bibliography{SubGE}

\begin{thebibliography}{}

\bibitem[\protect\citeauthoryear{Atchad{\'e} and Fort}{Atchad{\'e} and
  Fort}{2010}]{atchade2010limit}
Atchad{\'e}, Y. and G.~Fort (2010).
\newblock Limit theorems for some adaptive {MCMC} algorithms with subgeometric
  kernels.
\newblock {\em Bernoulli\/}~{\em 16}, 116--154.

\bibitem[\protect\citeauthoryear{Cline}{Cline}{2007}]{cline2007stability}
Cline, D. B.~H. (2007).
\newblock Stability of nonlinear stochastic recursions with application to
  nonlinear {AR}--{GARCH} models.
\newblock {\em Advances in Applied Probability\/}~{\em 39}, 462--491.

\bibitem[\protect\citeauthoryear{Cline and Pu}{Cline and
  Pu}{1998}]{cline1998verifying}
Cline, D. B.~H. and H.~H. Pu (1998).
\newblock Verifying irreducibility and continuity of a nonlinear time series.
\newblock {\em Statistics \& Probability Letters\/}~{\em 40}, 139--148.

\bibitem[\protect\citeauthoryear{Cline and Pu}{Cline and
  Pu}{2004}]{cline2004stability}
Cline, D. B.~H. and H.~H. Pu (2004).
\newblock Stability and the {L}yapounov exponent of threshold {AR--ARCH}
  models.
\newblock {\em Annals of Applied Probability\/}~{\em 14}, 1920--1949.

\bibitem[\protect\citeauthoryear{Davidson}{Davidson}{1994}]{Davidson1994}
Davidson, J. (1994).
\newblock {\em Stochastic Limit Theory}.
\newblock Oxford: Oxford University Press.

\bibitem[\protect\citeauthoryear{Douc, Fort, Moulines, and Soulier}{Douc
  et~al.}{2004}]{douc2004practical}
Douc, R., G.~Fort, E.~Moulines, and P.~Soulier (2004).
\newblock Practical drift conditions for subgeometric rates of convergence.
\newblock {\em Annals of Applied Probability\/}~{\em 14}, 1353--1377.

\bibitem[\protect\citeauthoryear{Douc, Guillin, and Moulines}{Douc
  et~al.}{2008}]{douc2008bounds}
Douc, R., A.~Guillin, and E.~Moulines (2008).
\newblock Bounds on regeneration times and limit theorems for subgeometric
  {M}arkov chains.
\newblock {\em Annales de l'Institut Henri Poincar{\'e} -- Probabilit{\'e}s et
  Statistiques\/}~{\em 44}, 239--257.

\bibitem[\protect\citeauthoryear{Douc, Moulines, Priouret, and Soulier}{Douc
  et~al.}{2018}]{douc2018markov}
Douc, R., E.~Moulines, P.~Priouret, and P.~Soulier (2018).
\newblock {\em Markov Chains}.
\newblock Cham: Springer.

\bibitem[\protect\citeauthoryear{Dudley}{Dudley}{2004}]{Dudley2004}
Dudley, R.~M. (2004).
\newblock {\em Real Analysis and Probability}.
\newblock Cambridge: Cambridge University Press.

\bibitem[\protect\citeauthoryear{Engle}{Engle}{1982}]{engle1982autoregressive}
Engle, R.~F. (1982).
\newblock Autoregressive conditional heteroscedasticity with estimates of the
  variance of {U}nited {K}ingdom inflation.
\newblock {\em Econometrica\/}~{\em 50}, 987--1007.

\bibitem[\protect\citeauthoryear{Fefferman and Shapiro}{Fefferman and
  Shapiro}{1972}]{FeffermanShapiro1972}
Fefferman, C. and H.~S. Shapiro (1972).
\newblock A planar face on the unit sphere of the multiplier space ${M}_p$,
  $1<p<\infty$.
\newblock {\em Proceedings of the American Mathematical Society\/}~{\em 36},
  435--439.

\bibitem[\protect\citeauthoryear{Fort and Moulines}{Fort and
  Moulines}{2000}]{fort2000vsubgeometric}
Fort, G. and E.~Moulines (2000).
\newblock V-subgeometric ergodicity for a {H}astings--{M}etropolis algorithm.
\newblock {\em Statistics \& Probability Letters\/}~{\em 49}, 401--410.

\bibitem[\protect\citeauthoryear{Fort and Moulines}{Fort and
  Moulines}{2003}]{fort2003polynomial}
Fort, G. and E.~Moulines (2003).
\newblock Polynomial ergodicity of {M}arkov transition kernels.
\newblock {\em Stochastic Processes and their Applications\/}~{\em 103},
  57--99.

\bibitem[\protect\citeauthoryear{Horn and Johnson}{Horn and
  Johnson}{2013}]{HornJohnson2013}
Horn, R.~A. and C.~R. Johnson (2013).
\newblock {\em Matrix Analysis\/} (2nd ed.).
\newblock Cambridge University Press.

\bibitem[\protect\citeauthoryear{Jarner and Roberts}{Jarner and
  Roberts}{2002}]{jarner2002polynomial}
Jarner, S.~F. and G.~O. Roberts (2002).
\newblock Polynomial convergence rates of {M}arkov chains.
\newblock {\em Annals of Applied Probability\/}~{\em 12}, 224--247.

\bibitem[\protect\citeauthoryear{Jones and Faddy}{Jones and
  Faddy}{2003}]{jones2003skew}
Jones, M.~C. and M.~J. Faddy (2003).
\newblock A skew extension of the $t$-distribution, with applications.
\newblock {\em Journal of the Royal Statistical Society: Series B\/}~{\em 65},
  159--174.

\bibitem[\protect\citeauthoryear{Klokov}{Klokov}{2007}]{klokov2007lower}
Klokov, S.~A. (2007).
\newblock Lower bounds of mixing rate for a class of {M}arkov processes.
\newblock {\em Theory of Probability and Its Applications\/}~{\em 51},
  528--535.

\bibitem[\protect\citeauthoryear{Klokov and Veretennikov}{Klokov and
  Veretennikov}{2004}]{klokov2004sub}
Klokov, S.~A. and A.~{\relax Yu}. Veretennikov (2004).
\newblock Sub-exponential mixing rate for a class of {M}arkov chains.
\newblock {\em Mathematical Communications\/}~{\em 9}, 9--26.

\bibitem[\protect\citeauthoryear{Klokov and Veretennikov}{Klokov and
  Veretennikov}{2005}]{klokov2005subexponential}
Klokov, S.~A. and A.~{\relax Yu}. Veretennikov (2005).
\newblock On subexponential mixing rate for {M}arkov processes.
\newblock {\em Theory of Probability and Its Applications\/}~{\em 49},
  110--122.

\bibitem[\protect\citeauthoryear{Lieberman and Phillips}{Lieberman and
  Phillips}{2020}]{lieberman2020hybrid}
Lieberman, O. and P.~C.~B. Phillips (2020).
\newblock Hybrid stochastic local unit roots.
\newblock {\em Journal of Econometrics\/}~{\em 215}, 257--285.

\bibitem[\protect\citeauthoryear{Ling}{Ling}{1999}]{ling1999probabilistic}
Ling, S. (1999).
\newblock On the probabilistic properties of a double threshold {ARMA}
  conditional heteroskedastic model.
\newblock {\em Journal of Applied Probability\/}~{\em 36}, 688--705.

\bibitem[\protect\citeauthoryear{Ling and McAleer}{Ling and
  McAleer}{2002}]{ling2002necessary}
Ling, S. and M.~McAleer (2002).
\newblock Necessary and sufficient moment conditions for the {GARCH}($r$,$s$)
  and asymmetric power {GARCH}($r$,$s$) models.
\newblock {\em Econometric Theory\/}~{\em 18}, 722--729.

\bibitem[\protect\citeauthoryear{Liu, Li, and Li}{Liu
  et~al.}{1997}]{liu1997threshold}
Liu, J., W.~K. Li, and C.~W. Li (1997).
\newblock On a threshold autoregression with conditional heteroscedastic
  variances.
\newblock {\em Journal of Statistical Planning and Inference\/}~{\em 62},
  279--300.

\bibitem[\protect\citeauthoryear{Meitz and Saikkonen}{Meitz and
  Saikkonen}{2008a}]{meitz2008ergodicity}
Meitz, M. and P.~Saikkonen (2008a).
\newblock Ergodicity, mixing, and existence of moments of a class of {M}arkov
  models with applications to {GARCH} and {ACD} models.
\newblock {\em Econometric Theory\/}~{\em 24}, 1291--1320.

\bibitem[\protect\citeauthoryear{Meitz and Saikkonen}{Meitz and
  Saikkonen}{2008b}]{meitz2008stability}
Meitz, M. and P.~Saikkonen (2008b).
\newblock Stability of nonlinear {AR--GARCH} models.
\newblock {\em Journal of Time Series Analysis\/}~{\em 29}, 453--475.

\bibitem[\protect\citeauthoryear{Meitz and Saikkonen}{Meitz and
  Saikkonen}{2010}]{meitz2010SPL}
Meitz, M. and P.~Saikkonen (2010).
\newblock A note on the geometric ergodicity of a nonlinear {AR--ARCH} model.
\newblock {\em Statistics \& Probability Letters\/}~{\em 80}, 631--638.

\bibitem[\protect\citeauthoryear{Meitz and Saikkonen}{Meitz and
  Saikkonen}{2021}]{meitz2021JAP}
Meitz, M. and P.~Saikkonen (2021).
\newblock Subgeometric ergodicity and $\beta$-mixing.
\newblock {\em Journal of Applied Probability\/}~{\em 58}, 594--608.

\bibitem[\protect\citeauthoryear{Meitz and Saikkonen}{Meitz and
  Saikkonen}{2022}]{meitz2022subgear}
Meitz, M. and P.~Saikkonen (2022).
\newblock Subgeometrically ergodic autoregressions.
\newblock {\em Econometric Theory\/}~{\em 38}, 959--985.

\bibitem[\protect\citeauthoryear{Merlev{\`e}de, Peligrad, and
  Rio}{Merlev{\`e}de et~al.}{2011}]{merlevede2011bernstein}
Merlev{\`e}de, F., M.~Peligrad, and E.~Rio (2011).
\newblock A {B}ernstein type inequality and moderate deviations for weakly
  dependent sequences.
\newblock {\em Probability Theory and Related Fields\/}~{\em 151}, 435--474.

\bibitem[\protect\citeauthoryear{Meyn and Tweedie}{Meyn and
  Tweedie}{2009}]{meyn2009markov}
Meyn, S.~P. and R.~L. Tweedie (2009).
\newblock {\em Markov Chains and Stochastic Stability\/} (2nd ed.).
\newblock Cambridge: Cambridge University Press.

\bibitem[\protect\citeauthoryear{Nummelin and Tuominen}{Nummelin and
  Tuominen}{1983}]{nummelin1983rate}
Nummelin, E. and P.~Tuominen (1983).
\newblock The rate of convergence in {O}rey's theorem for {H}arris recurrent
  {M}arkov chains with applications to renewal theory.
\newblock {\em Stochastic Processes and their Applications\/}~{\em 15},
  295--311.

\bibitem[\protect\citeauthoryear{Phillips}{Phillips}{2023}]{phillips2023estimation}
Phillips, P. C.~B. (2023).
\newblock Estimation and inference with near unit roots.
\newblock {\em Econometric Theory\/}~{\em 39}, 221--263.

\bibitem[\protect\citeauthoryear{Tuominen and Tweedie}{Tuominen and
  Tweedie}{1994}]{tuominen1994subgeometric}
Tuominen, P. and R.~L. Tweedie (1994).
\newblock Subgeometric rates of convergence of $f$-ergodic {M}arkov chains.
\newblock {\em Advances in Applied Probability\/}~{\em 26}, 775--798.

\bibitem[\protect\citeauthoryear{Tweedie}{Tweedie}{1983}]{tweedie1983criteria}
Tweedie, R.~L. (1983).
\newblock Criteria for rates of convergence of {M}arkov chains, with
  application to queueing and storage theory.
\newblock In J.~F.~C. Kingman and G.~E.~H. Reuter (Eds.), {\em Probability,
  Statistics and Analysis}, pp.\  260--276. Cambridge: Cambridge University
  Press.

\bibitem[\protect\citeauthoryear{Veretennikov}{Veretennikov}{2000}]{veretennikov2000polynomial}
Veretennikov, A.~{\relax Yu}. (2000).
\newblock On polynomial mixing and convergence rate for stochastic difference
  and differential equations.
\newblock {\em Theory of Probability and Its Applications\/}~{\em 44},
  361--374.

\bibitem[\protect\citeauthoryear{Vladimirova, Girard, Nguyen, and
  Arbel}{Vladimirova et~al.}{2020}]{vladimirova2020sub}
Vladimirova, M., S.~Girard, H.~Nguyen, and J.~Arbel (2020).
\newblock Sub-{W}eibull distributions: {G}eneralizing sub-{G}aussian and
  sub-{E}xponential properties to heavier tailed distributions.
\newblock {\em Stat\/}~{\em 9}, e318.

\bibitem[\protect\citeauthoryear{Wong, Li, and Tewari}{Wong
  et~al.}{2020}]{wong2020lasso}
Wong, K.~C., Z.~Li, and A.~Tewari (2020).
\newblock Lasso guarantees for $\beta$-mixing heavy-tailed time series.
\newblock {\em Annals of Statistics\/}~{\em 48}, 1124--1142.

\end{thebibliography}

\end{document}